\documentclass[10pt,conference]{IEEEtran}
\IEEEoverridecommandlockouts
\usepackage{cite}
\usepackage{hyperref}
\hypersetup{
    colorlinks=true,
    linkcolor=blue,
    filecolor=blue,      
    urlcolor=blue,
    pdfpagemode=FullScreen
    }
\usepackage{booktabs}
\usepackage{amsmath,amssymb,amsfonts}

\usepackage{algorithmic}
\usepackage{graphicx}
\usepackage{textcomp}
\def\BibTeX{{\rm B\kern-.05em{\sc i\kern-.025em b}\kern-.08em
    T\kern-.1667em\lower.7ex\hbox{E}\kern-.125emX}}
\usepackage{enumitem}
\usepackage{mathtools}
\usepackage{pifont}
\usepackage[font=footnotesize]{caption}
\usepackage{subcaption}
\usepackage[linesnumbered,ruled,vlined]{algorithm2e}
\usepackage{numprint}
\usepackage{makecell}
\usepackage{tikz}
\usepackage{orcidlink}

\newcommand{\systemname}{\textsc{FCPO}}
\newcommand{\iagent}{\textit{iAgent}}
\newcommand{\bcheckmark}{\ding{51}} % Bold checkmark
\newcommand{\btimes}{\ding{55}}     % Bold times (cross) symbol

\usepackage{xcolor}

\newcommand{\sourcecodelink}{https://github.com/tungngreen/PipelineScheduler}
\newcommand{\archivelink}{https://doi.org/10.5281/zenodo.14789255}

\definecolor{stepcolor}{HTML}{990000}
\definecolor{stepfill}{HTML}{FFA500}

\newcommand*\circled[1]{\tikz[baseline=(char.base)]{
            \node[shape=circle,draw,inner sep=2pt] (char) {{\textbf{#1}}};}}

\definecolor{applegreen}{rgb}{0.55, 0.71, 0.0}

\definecolor{caribbeanblue}{rgb}{0.05, 0.9, 0.9}

\begin{document}

\title{\systemname: Federated Continual Policy Optimization for Real-Time High-Throughput Edge Video Analytics}

\author{
    \IEEEauthorblockN{1\textsuperscript{st} Lucas Liebe \orcidlink{0009-0004-9252-4764}, 1\textsuperscript{st} Thanh-Tung Nguyen \orcidlink{0000-0002-8186-2600}, 3\textsuperscript{rd} Dongman Lee}
    \IEEEauthorblockA{\textit{School of Computing}, \textit{KAIST}, Daejeon, Republic of Korea \\
    \{lucasliebe, tungnt, dlee\}@kaist.ac.kr}
}

\maketitle

\begin{abstract}
The growing complexity of Edge Video Analytics (EVA) facilitates new kind of intelligent applications, but creates challenges in real-time inference serving systems.
State-of-the-art (SOTA) scheduling systems optimize global workload distributions for heterogeneous devices but often suffer from extended scheduling cycles, leading to sub-optimal processing in rapidly changing Edge environments.
Local Reinforcement Learning (RL) enables quick adjustments between cycles but faces scalability, knowledge integration, and adaptability issues.
Thus, we propose \systemname{}, which combines Continual RL (CRL) with Federated RL (FRL) to address these challenges.
This integration dynamically adjusts inference batch sizes, input resolutions, and multi-threading during pre- and post-processing.
CRL allows agents to learn from changing Markov Decision Processes, capturing dynamic environmental variations, while FRL improves generalization and convergence speed by integrating experiences across inference models.
\systemname{} combines these via an agent-specific aggregation scheme and a diversity-aware experience buffer.
Experiments on a real-world EVA testbed showed over $5\times$ improvement in effective throughput, $60\%$ reduced latency, and $20\%$ faster convergence with up to $10\times$ less memory consumption compared to SOTA RL-based approaches.
\end{abstract}


\section{Introduction}
\label{sec:introduction}

Video Analytics (VA) is widely regarded as a "killer application" in Edge Computing~\cite{Ananthanarayanan2017vakiller}.
High-demand applications 
%in Edge Video Analytics (VA)
, such as traffic monitoring~\cite{Nguyen2023preacto} and  surveillance~\cite{Hung2018VideoEdge}, generate substantial data volumes that require local processing to enhance data privacy, minimize network latency and improve throughput~\cite{nguyen2025octopinf}.
VA services are typically organized into a series of tasks as pipelines.
To optimize latency and throughput of the whole pipeline,
approaches such as \cite{zeng2020distream, Hou2023dystri, ma2024performance, gao2024energy, nguyen2025octopinf} periodically perform scheduling to balance the workloads assigned to the edge server and devices, avoiding performance bottlenecks.

\begin{figure}[t]
    \centering
    \includegraphics[width=0.85\linewidth, trim=10 10 10 0, clip]{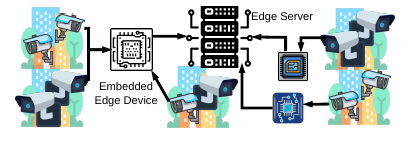}
    \caption{A VA scenario with collaborating Edge devices. \textit{Various sizes, shapes, and colors illustrate architecture, resource, and workload heterogeneity.}}
    \label{fig:edge-va-intro}
    % \vspace{-1.5em}
\end{figure}

Once assigned, each device performs its assigned tasks until the next scheduling period, ranging from a few minutes~\cite{nguyen2025octopinf} to 3 hrs~\cite{Hung2018VideoEdge}.
%We identify that dynamic real-time edge applications require continual performance optimization at the device level during this period, due to factors such as fluctuating network conditions and workload variability.
%These dynamics are further compounded through complexity by the heterogeneity of computing devices (\autoref{fig:edge-va-intro}), which span a range of x86-64 servers and arch64 embedded computers (e.g., NVIDIA Jetsons, Google Corals, and Rockchips) with diverse computational capacities and resource availability~\cite{chen2024scenic}.
We identify that dynamic real-time edge applications require continual performance optimization at the device level during this period due to multiple factors.
These factors include fluctuating network conditions, workload variability, and the heterogeneity of computing devices (\autoref{fig:edge-va-intro}), spanning a range of x86-64 servers and arch64 embedded computers with diverse computational capacities and resource availability~\cite{nguyen2025octopinf}.
Thus, on-device fast-paced and constant local optimization is essential to meet real-time requirements, as their violations can cause catastrophic failures and unsafe behavior~\cite{chen2024scenic}.

To this end, several studies have proposed \textit{Reinforcement Learning (RL)}-based approaches, adaptively learning near-optimal solutions at the edge device~\cite{zhang2024bcedge, fang2017qos,she2024earlyexit,coviello2021magicpipe}.
However, they have yet to overcome four major limitations.

\noindent \textbf{\textit{(1) Offline training.}} Although RL can learn in real-time from newly obtained experiences, this capability is often limited by computational complexity (particularly in Deep RL) and concerns over reliability and stability~\cite{osinenko2023actor}.
For example, QoSAS \cite{fang2017qos}, BCEdge~\cite{zhang2024bcedge}, and DDQN~\cite{she2024earlyexit} rely on offline-trained RL agents, utilizing only the inference phase at runtime.
This approach restricts the real-time adaptability of RL, and these pre-trained agents may experience performance degradation, due to large environmental variations at the Edge.

\noindent \textbf{\textit{(2) Poor Scalability.}} In existing approaches~\cite{fang2017qos, zhang2024bcedge, she2024earlyexit}\footnotemark[1], adaptation is achieved by deploying one RL agent per device.
Since these agents rely on offline training, they must continually collect and store experiences as data for retraining.
As the number of inference models increases, the storage and time required for training also grow, with each iteration processing more data samples.
This rapidly becomes a scalability bottleneck, restricting efficiency in large-scale deployments.

\footnotetext[1]{\label{fnote:magicpipe}MagicPipe \cite{coviello2021magicpipe} does not provide an algorithm on how to scale up when new devices or models are introduced into the cluster.}

\noindent \textbf{\textit{(3) 
Learning Divergence.}} In current approaches~\cite{fang2017qos, zhang2024bcedge, she2024earlyexit}\footnotemark[1], when a new environment is encountered, a base agent is typically cloned and then the clone is tasked with optimization for that environment.
Although these agents begin with similar initial understanding, their learning diverges over time due to \textit{experience heterogeneity}.
This makes it difficult for these agents to have a consistent unified policy across collaborating edge devices and the server.

\noindent \textbf{\textit{(4) Cold start.}} 
Newly cloned agents often face significant cold start challenges when they are deployed in a new environment. 
These challenges occur because the agents are initialized with pre-trained parameters or policies from a previous setting. 
The severity of these cold start issues is closely related to the differences between the old and new environments.
When there are substantial disparities in state distributions, action dynamics, or reward structures, agents typically experience a significant decline in performance. 
This drop in performance results from the mismatch between the agent's learned representations and the unfamiliar conditions it encounters in the new environment.

\par
\textbf{System Overview and Contributions}.
To address these limitations of the existing works, we propose \systemname{} -- a system based on real-time \textit{Federated Continual Reinforcement Learning} for \textit{Policy Optimization} of pipeline configuration in latency-constrained EVA.
In \systemname{}, each lightweight \textit{inference agent (iAgent)} is specifically designed to manage and optimize a DNN inference model, rather than an entire device, fostering scalability even as the number of models grows.
Each iAgent adapts dynamically to heterogeneous environments by selecting among three actions: (a) inference batch size selection, which balances throughput and latency; (b) input resolution selection, optimizing between accuracy and computational load; and (c) multi-thread processing, enabling efficient parallelism based on current resource availability.
These actions represent complex and context-aware resource allocation strategies, allowing iAgents to continually adjust to changing conditions and workloads in real-time.
Our iAgent-based approach allows for finer-grained control over resource allocations and enhances adaptability across varying models and workloads.
To overcome the other limitations, we propose a novel learning method, \textit{Federated Continual Reinforcement Learning (FCRL)}, which integrates \textit{Continual Reinforcement Learning (CRL)} with \textit{Federated Reinforcement Learning (FRL)}.
Recent work has provided a foundational framework for CRL, defining a stable approach for continually training agents in dynamic environments~\cite{abel2024continual}.
Building on this, we incorporate FRL~\cite{Jin2022federatedreinforcementlearning} to facilitate collaborative learning across agents in heterogeneous, non-IID environments with varying state transitions.
To our knowledge, this is the first implementation of CRL combined with FRL in the novel FCRL framework for real-time systems.

\begin{table}[t]
\vspace{0.1em}
\centering
\caption{Comparisons to the state-of-the-art RL-based systems}
\vspace{-0.5em}
    \label{Tab:sota-comparison}
    \begin{tabular}{@{\hspace{1pt}} >{\centering\arraybackslash}p{1.8cm} 
                    >{\centering\arraybackslash}p{1.5cm}
                    @{\hspace{6pt}} >{\centering\arraybackslash}p{0.95cm}
                    @{\hspace{8pt}} >{\centering\arraybackslash}p{1.8cm}
                    @{\hspace{8pt}} >{\centering\arraybackslash}p{1cm}}
      \hline
      \toprule
      %\textbf{System} & \makecell{\textbf{Continual} \\ 
      %%%%%% Column titles
      \textbf{System} &
      \makecell{\textbf{Online} \\ \textbf{Learning}} & 
      \makecell{\textbf{Scala-} \\ \textbf{bility}} &
      \makecell{\textbf{Knowledge} \\ \textbf{Fusion}} &
      \makecell{\textbf{Warm} \\ \textbf{Start}} \\\hline

      %%%%%% Contents
      QoSAS~\cite{fang2017qos} & \btimes & \btimes & \btimes & \btimes \\\hline
      BCEdge~\cite{zhang2024bcedge} & \btimes & \btimes & \btimes & \textit{Last$^\ddagger$} \\\hline
      MagicPipe~\cite{coviello2021magicpipe} & \bcheckmark & \btimes & \btimes   & \btimes \\\hline
      DDQN~\cite{she2024earlyexit} & \btimes & \btimes & \btimes & \textit{Last$^\ddagger$} \\\hline
      \textbf{FCPO} & \bcheckmark & \bcheckmark & \bcheckmark & \bcheckmark \\\hline
    \end{tabular}
    \begin{flushleft}
        \small $^\ddagger$ Last training checkpoint.
    \end{flushleft}
    \vspace{-0.5em}
\end{table}

\textbf{Overall}, our contributions and benefits to existing works are shown in \autoref{Tab:sota-comparison} and summarized as follows:
\begin{itemize} [leftmargin=*]
    \item We propose \systemname{}, a scalable system for optimizing real-time performance of VA inference pipelines in large-scale edge systems by enabling fine-grained control over batch size, resolution, and multi-thread processing.
    \item We introduce FCRL, combining Federated and Continual Reinforcement Learning, to enable adaptive, collaborative learning in dynamic, non-iid edge environments for VA.
    \item We develop an agent-specific aggregation scheme that combines shared backbone knowledge with heterogeneous actions, handling large action spaces and environment-specific optimizations efficiently.
    \item We conduct real-world experiments, which show over $5\times$ improvement in effective throughput, $60\%$ reduced latency, and $20\%$ faster convergence with up to $10\times$ less memory consumption, demonstrating \systemname{}’s superiority in managing complex, large-scale VA tasks with enhanced adaptability compared to SOTA RL-based approaches.
\end{itemize}

\section{Preliminaries and Motivations}
\label{sec:system-design}

\subsection{Design Goals}
\label{subsec:design-goals}

\par
Similar to prior works~\cite{fang2017qos, coviello2021magicpipe, she2024earlyexit, zhang2024bcedge}, \systemname{} aims to achieve high throughput and low latency to ensure efficient, real-time processing of VA tasks.
In addition to these objectives, we emphasize minimizing convergence speed, decision latency, training latency, and optimizing resource utilization.
Real-time decision-making and adaptive learning processes can impose significant demands on hardware resources, especially in constrained edge environments. Therefore, \systemname{} is designed to rapidly converge on optimal configurations with minimal decision and training latency, while keeping resource usage as low as possible to maximize the available capacity for workloads.
This approach ensures that resources are primarily dedicated to VA processing rather than overhead, achieving both efficiency and responsiveness in large-scale deployments.

\subsection{Action Space for Quick Adaptation}
\label{subsec:action-space}

\begin{figure}[t]
    \centering
    \begin{subfigure}{0.44\linewidth}
        \centering
        \includegraphics[width=1.05\textwidth, trim=20 10 20 10, clip]{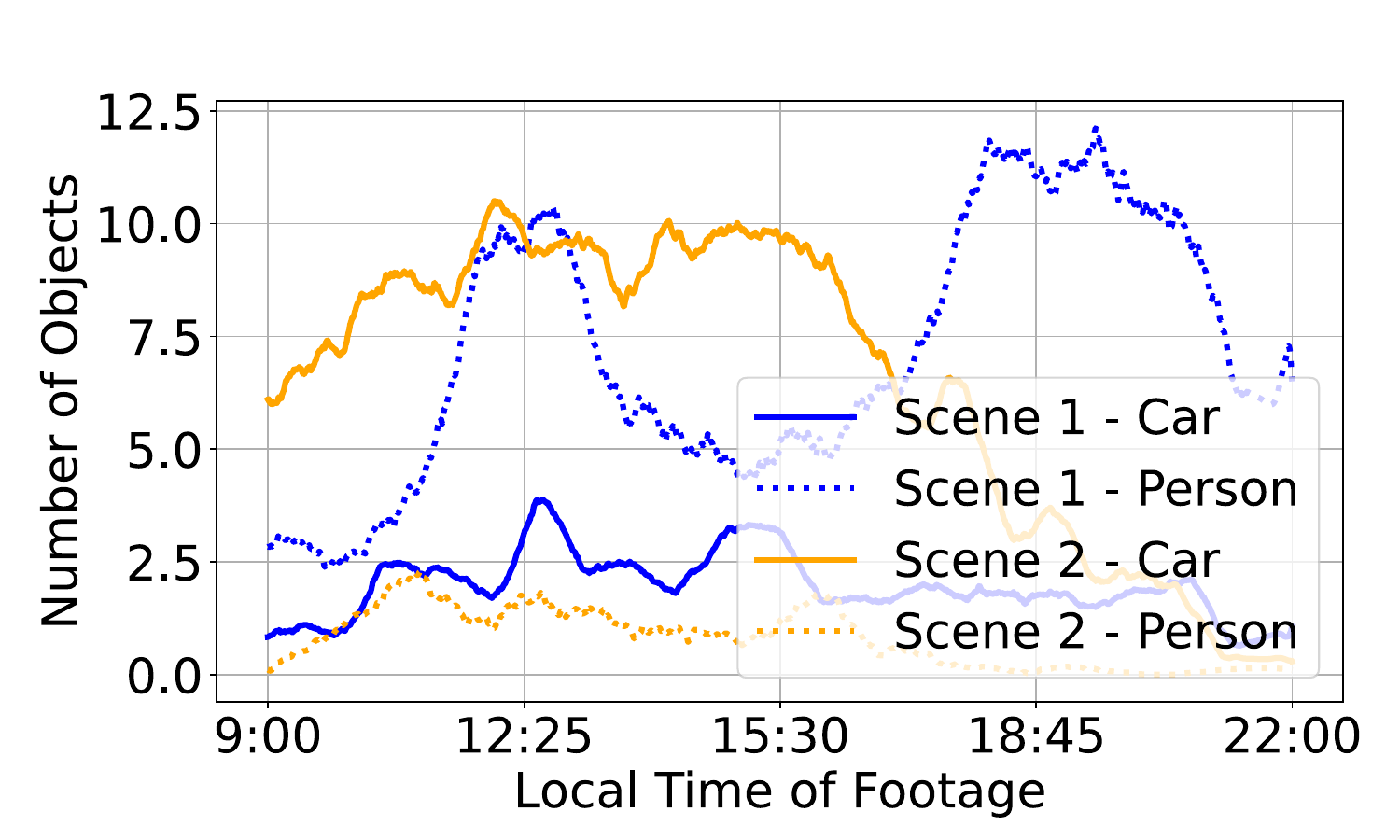}
        \label{fig:content-dynamics}    
        \vspace{-1.5em}
        \caption{Content dynamics in 2 traffic scenes}
    \end{subfigure}
    \quad
    \begin{subfigure}{0.45\linewidth}
        \centering
        \includegraphics[width=1.05\textwidth, trim=10 10 20 10, clip]{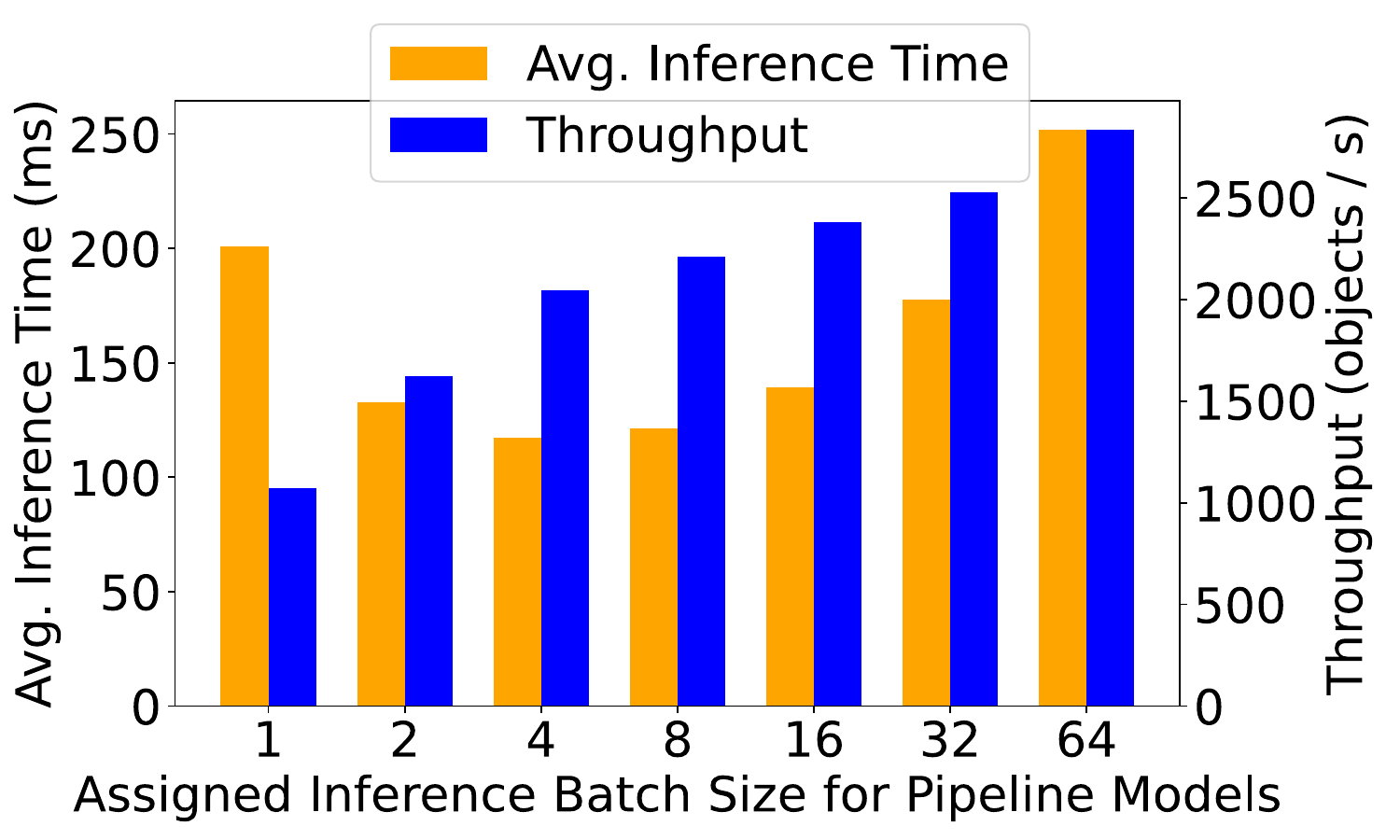}
        \label{fig:batching-throughput}   
        \vspace{-1.5em}
        \caption{Batch latency/throughput trade-off}
    \end{subfigure}
    \caption{Motivation for adaptation and dynamic batching.}
    \label{fig:batching-dynamics}
    % \vspace{-1em}
\end{figure}

\par
In the highly dynamic edge environment, fluctuating network conditions and variable workloads, as shown in \autoref{fig:batching-dynamics}a, introduce significant challenges to maintaining consistent performance.
These factors necessitate continual optimization to adapt to rapid changes in real-time as they can lead to degraded throughput, increased latency, and inefficient resource use.
To this end, we consider three methods with different throughput trade-offs as actions for the agent, which are described in the next sections.

\textbf{Dynamic Batched Inference.}
Batched inference has long been regarded an effective method to increase the inference throughput of DNN models~\cite{ali2020batch}, as it efficiently leverages the parallel capabilities of GPUs. 
By dynamically adjusting the batch size, it is possible to achieve desirable throughput levels. 
\autoref{fig:batching-dynamics}b demonstrates the impact of batching on throughput and end-to-end latency within an EVA traffic monitoring pipeline, highlighting two key insights:
(1) Larger batch sizes improve throughput via vector-level parallelism but increase latency, as the first query in a batch must wait for the last.
(2) Smaller batch sizes reduce individual inference latency but can create pipeline bottlenecks, causing higher end-to-end latency, as seen with batch sizes of 1 and 2.
Thus, while dynamic batching can be an effective tool for balancing throughput and latency, careful tuning is crucial for optimization.% to optimize the overall performance.

\textbf{Resolution Adjustments.}
Resizing input data (e.g., video frames) to improve throughput has been widely studied~\cite{nigade2022jellyfish,gokarn2023mosaic,peng2024tangram}.
However, traditional DNNs often lack support for variable input sizes, and using multiple inference engines can lead to excessive resource demands. %be resource-intensive in constrained environments.
\systemname{} addresses this with frame packing, combining smaller images into a single frame for inference models.
A similar method is proposed by Peng et al.~\cite{peng2024tangram} and Gokarn et al.~\cite{gokarn2023mosaic}.

\textbf{Multi-thread Processing.}
DNN-based query processing comprises three steps: (1) pre-processing (e.g., normalization), (2) DNN inference, and (3) post-processing (e.g., decoding and filtering).
While pre- and post-processing cannot be batched, throughput can be improved via concurrency, using multiple threads to prevent bottlenecks.
However, on resource-limited embedded devices, excessive threading may degrade performance due to resource contention.
Carefully managing thread allocation is crucial to balance concurrency benefits with resource efficiency, avoiding system overload.

\subsection{Challenges for Continual Adaptation in Real-Time}

Continual learning of RL agents in a real-world environment, rather than relying on simulations, is critical due to the inherent variability of edge scenarios.
Simulations often fail to capture the whole stochastic nature of network conditions, hardware heterogeneity, and workload variability.
However, achieving continual learning with real-time training presents four critical challenges not yet overcome by existing works:

% \begin{enumerate} [leftmargin=*]
%     \item \textit{Exponential Action Combinations}: The combination of actions required for optimal performance in edge environments creates an exponential search space, making naive random exploration infeasible. In \systemname{}, the agent must dynamically adjust the batch size, resolution, and multi-threading level. Selecting an optimal batch size while simultaneously tuning frame packing and thread counts results in a vast array of configurations, each affecting latency, throughput, and resource usage. Efficient exploration strategies are thus essential to identify the best action combinations without excessive sampling.
%     \item \textit{Learning Overhead}: Real-time learning introduces computational overhead which can create contention for critical resources such as CPU cycles, memory bandwidth, and GPU processing power, potentially leading to slower inference times and degraded performance.
%     \item \textit{Non-IID Experiences}: Each agent’s experience is unique, as it encounters distinct sequences of states and actions due to varied exploration paths, workload patterns, and environmental conditions. This non-iid nature complicates knowledge aggregation and fusion across agents, as they evolve based on distinct data distributions.
%     \item \textit{Heterogeneous Action Spaces}: Due to device and model heterogeneity, resource constraints may limit viable actions (e.g., batch size or resolution choices), making the action spaces inconsistent across agents.
% \end{enumerate}

\subsubsection{\textbf{Exponential Action Combinations}} The combination of actions required for optimal performance in edge environments creates an exponential search space, making naive random exploration infeasible.
In \systemname{}, the agent must dynamically adjust the batch size, resolution, and multi-threading level.
Selecting the optimal batch while simultaneously tuning frame packing and thread counts results in a vast array of configurations, each affecting latency, throughput, and resource usage. Efficient exploration strategies are thus essential to identify the best action combinations without excessive sampling.
\subsubsection{\textbf{Learning Overhead}} Real-time learning introduces computational overhead which can create contention for critical resources such as CPU cycles, memory bandwidth, and GPU processing power, potentially leading to slower inference times and degraded performance.
\subsubsection{\textbf{Non-IID Experiences}} Each agent’s experience is unique, as it encounters distinct sequences of states and actions due to varied exploration paths, workload patterns, and environmental conditions. This non-iid nature complicates knowledge aggregation and fusion across agents, as they evolve based on distinct data distributions.
\subsubsection{\textbf{Heterogeneous Action Spaces}} Due to device and model heterogeneity, resource constraints may limit viable actions (e.g., batch size or resolution choices), making the action spaces inconsistent across agents.

The next section covers our approach to address these challenges, which is essential to ensure that RL agents can effectively learn in real-world edge systems, maintaining performance and reliability across diverse and variable conditions.

\section{Real-time Inference Serving System}
\label{sec:system}

\subsection{System Overview}
\label{subsec:overview}

\autoref{fig:components} shows the components for controlling the real-time edge VA system.
\systemname{} follows a popular setup for edge computing systems, consisting of multiple clusters~\cite{zeng2020distream}.
Each cluster consists of a local server and multiple heterogeneous edge devices, connected to various data sources (e.g., real-time cameras).
The local server runs the \textit{System Controller} responsible for model allocation, system-wide scheduling, and executing \textit{Agent-Specific FL Aggregation} (\autoref{subsec:method_federated_optimization}).
In this paper, we focus on perform continual local real-time optimization and leverage ~\cite{nguyen2025octopinf} for global periodic scheduling.

Both the server and devices are capable of hosting VA inference models.
Due to their heterogeneous resource availability, the number of models hosted at each machine varies significantly.
Each host device is controlled by a single \textit{Device Control}, responsible for running the models according to the scheduled provided by \textit{System Controller} and collecting device run-time statistics.
In BCEdge~\cite{zhang2024bcedge} this component contains the intelligent agent, while in \systemname{}, this component is responsible for real-time metrics collection and FL participation.
Communication from \textit{System Controller} to the workload is coordinated through the \textit{Device Control}, ensuring fair and sequential execution of system changes.

Each \textit{Workload Model} is piggybacked  with a light-weight \textit{Continual RL (CRL)} agent, called \iagent{}, in charge of organizing the structure between processing threads and collects workload metrics.
%In \systemname{} this entity contains a light-weight \textit{Continual RL (CRL)} agent that is called \iagent{}.
\iagent{} constantly observes the model's performance against its environment and adaptively learns the optimal configuration to improve the performance.

The final component in the system is a distributed \textit{Metric Database (DB)} for storing real-time metrics that are used to evaluate the system and for updating allocation strategies.

\subsection{System Procedure}
\label{subsec:workflow}

\begin{figure}[t]
    \centering
    \begin{subfigure}{\linewidth}
        \includegraphics[width=\textwidth]{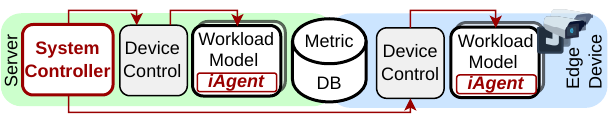}
        \vspace{-2em}
        \caption{Overview of the main components.}
        \vspace{1em}
        \label{fig:components}
    \end{subfigure}
    \begin{subfigure}{\linewidth}
        \includegraphics[width=\textwidth]{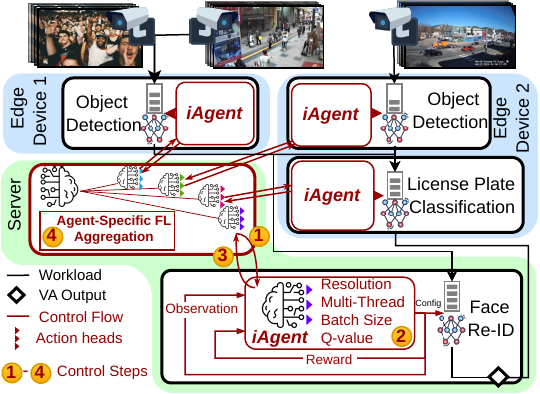}
        \caption{Detailed workflow and architecture for \systemname{}'s real-time pipeline processing and \iagent{} Federated Continual Reinforcement learning (FCRL).}
        \label{fig:workflow}
    \end{subfigure}
    \caption{System-wide \systemname{} architecture overview.}
    % \vspace{-1.5em}
\end{figure}

\autoref{fig:workflow} illustrates the architecture of \systemname{}, a system based on \textit{Federated Continual Reinforcement Learning (FCRL)} for continual Policy Optimization in real-world edge environments.
The workflow of the components in \systemname{} follows a typical FL procedure.

\begin{enumerate}[label=\protect\circled{\arabic*}, wide]
    \item At each round's start, the \textit{System Controller} distributes the aggregated \textit{Global Model (GM)} to all \iagent{}s. This helps \iagent{}s optimize their respective inference models, particularly assisting newly deployed models that might otherwise face cold starts.
    \item Each \iagent{} interacts with its specific environment using CRL to efficiency adapt itself to the environment.  
    \item Once local training is complete, selected \iagent{}s upload the adapted model parameters to the server.
    \item The \textit{System Controller} aggregates the local updates into a new GM. As introduced in \autoref{sec:introduction}, the knowledge learned by one agent can greatly benefit the others.
    \systemname{} employs Federated Learning to combine the collective knowledge learned by agents. 
    To avoid overgeneralization causing each agent to lose the environment-specific knowledge, \systemname{} is equipped with an \textit{Agent-Specific FL Aggregation} algorithm.
\end{enumerate}

\par
In the next section, we go into the details of \systemname{} by first introducing the architecture of \iagent{} in \autoref{subsec:iagent} and how its learning can be formalized into a Markov Decision Process (MDP) in \autoref{subsec:method_mdp}. 
Then we introduce the novel FCRL method to allow effective continual adaptation in \autoref{subsec:method_continual_learning} and \ref{subsec:method_federated_optimization}.

\section{Federated Continual Policy Optimization}
\label{sec:fcpo}

\subsection{\iagent's Model Architecture}
\label{subsec:iagent}
The model architecture of \iagent{} is shown in \autoref{fig:iagent-arch}.
It takes in an input of size 8, which includes the current actions, the arrival rate, and the number of drops from the full queue.
It comprises a backbone, one value head, and three action heads.
The backbone is implemented with two linear layers, featuring a hidden dimension of 64 and an output dimension of 48. 
These layers serve as a feature extractor for the subsequent layers, aiming to capture the overall dynamics of the environment.
The value head is designed as a single linear layer to estimate the cumulative reward.

\par
While all three actions are closely related, choosing all three actions together using a single action head results in a large exploration space for this head and consequently slower convergence.
Thus, to generate the probability distribution for action sampling, the network has one linear layer for each action, followed by a softmax activation function.
To facilitate feature sharing and alternating optimization among the heads, we take inspiration from Faster R-CNN model architecture with cascading outputs~\cite{mansour2021rcnn}.

\par
Particularly, the first action head utilizes the backbone's features to determine the resolution. 
Its output is then concatenated into the backbone features to determine the actions of the other two heads.
This approach enables the agent to learn the dependencies among the actions. 
For instance, if the request rate is static and the resolution is lowered to accommodate two images in a frame, the throughput is doubled and the batch size can be reduced.
Additionally, lowering the resolution increases the preprocessing rate, resulting in the necessity for an additional preprocessing thread to avoid bottlenecks.

\subsection{Markov Decision Process of \iagent{}}
\label{subsec:method_mdp}
\begin{figure}[t]
    \centering
    \includegraphics[width=\linewidth]{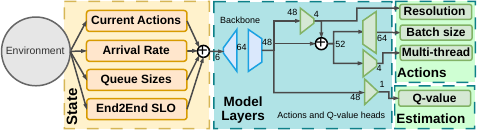}
    \vspace{-2em}
    \caption{Architecture of \iagent.}
    \label{fig:iagent-arch}
    % \vspace{-1.9em}
\end{figure}

We describe the MDP of \iagent{} as consisting of states $S$, actions $A$, a transition probability distribution $P$, and a reward function $R$.
Every step $n \in \mathbb{N}$, \iagent{} collects the state and chooses an action.
Multiple steps within an episode are used to optimize the behavior in the next episode, as shown in \autoref{fig:iagent-arch} and \autoref{fig:model_architecture}.
\iagent{} learns the environment defined through the MDP and identifies an optimal policy $\pi^* = \text{argmax}_a Q^*(s,a)_n$, that maximizes the cumulative expected reward for a state-action-pair at step $n$.
The optimal Q-function $Q^*(s,a)_n$ defines the reward by choosing action $a\in A$ in state $s\in S$, which is approximated through a neural network.
% and configured with discount factor $\gamma$ to balance between the immediate and future rewards:
% \begin{equation}
%     Q^*(s,a)_n = \mathbb{E}_{\gamma \sim p(s_{n-1})}[r_n + \gamma \max_{a_{n-1}} Q^*(s, a)_{n-1}]
% \end{equation}

\textbf{\textit{State Space $S$}.}
A state space captures the characteristics of an environment. 
At the beginning of a step, the agent constructs a state $s_n \in S$ as a vector of size 8, $s_n \in S \subset \mathbb{R}^8$.
The most important state metric is the request rate, as it determines the throughput the agent needs to achieve.
Other metrics guide the agent in its decision-making process, such as the current resolution configuration, batch size, and thread configuration.
The current queue sizes between processing steps are also incorporated to help the agent identify processing bottlenecks.
While these inputs may not be strictly necessary for a well-trained agent, they can significantly improve the learning speed, which is crucial for online-training.
The last state value is the end-to-end \textit{Service Level Objectives (SLO)} of the pipeline, it is used to give the agent context for what kind of processing speed is required.

\textbf{\textit{Action Space $A$}.}
An action refers to the behavior chosen by the agent for the current step $n$.
For \iagent{}, each action $a_n\in A \subset \mathbb{N}^3 $ is represented by a three-tuple:
\[
    [RES (\text{resolution}),~BS (\text{batch size}),~MT (\text{multi-thread})]
\]
All actions can be changed instantly at runtime, and have a direct impact on inference throughput. 
The optimal choice is to increase $a_n[1] = BS$, as decreasing the resolution may decrease accuracy, and adding threads allocates additional resources. 
Nevertheless, $RES$ and $MT$ may be necessary to handle a high workload with low latency.
%To encourage the agent to prioritize increasing the batch size, a small penalty will be applied for larger values of $a_n[0]$ and $a_n[2]$.

\textbf{\textit{Transition Probability Distribution $P$}.}
A transition probability $p_{s_{n+1}}\in P$ defines how likely an agent will transition from current state $s_n$ to the next state $s_{n+1}$ based on the chosen action $a_n$.
$p_{s_{n+1}} = P(s_{n-1}|s_n,a_n)$ captures the environment dynamics as well as the effects of an action.

\textbf{\textit{Reward Function $R$}.}
The reward function returns a scalar value $r_n$, rating the last action, and is designed to directly reflect the design goals by subtracting latency and an oversize penalty from the throughput.
These values are aggregated as the cumulated expected reward $\mathbb{E}[\sum^{|C|}_{n=0}\gamma^n r_n]$, where $r_n$ is the reward calculated after step $n$.
The reward is normalized between -1 and 1, resulting in the following equation for $R$ with $lat$ as the estimated weighted average of the local latency: 
\begin{equation}
\label{eq:reward-function}
        r_n = \frac{1}{2} * (\vartheta\frac{Throughput_n}{RequestRate_n} - \varsigma~lat - \varphi\frac{a[1]_n}{RequestRate_n})
\end{equation}
Compared to BCEdge~\cite{zhang2024bcedge} we do not directly include the model SLO into the reward function, because (a) \iagent{} should learn the relationship of throughput and latency and (b) each models deadline within a pipeline is ambiguous.
However, the oversize penalty is increased by the number of requests that exceed the local SLO. 
This way the reward is indirectly decreased for not meeting SLOs.

Finally, the \textit{\textbf{optimal reward function}} can be written as:
\begin{equation}
\label{eq:q-function}
    Q^*(s,a)_n = \mathbb{E}_{\gamma \sim p(s_{n-1})}[r_n + \gamma \max_{a_{n-1}} Q^*(s, a)_{n-1}]
\end{equation}
where $\gamma$ is the discount factor to balance between the immediate and future rewards.

\subsection{Continual Reinforcement Learning (CRL)}
\label{subsec:method_continual_learning}

As long as dynamic changes follow a consistent pattern, they can be represented within a single MDP.
For example, over short intervals, unstable network bandwidth can be captured by a single probability distribution \cite{Lubben2011networkmdp}.
However, averaging this distribution over longer periods may lead to imprecision, as edge deployments often face varying patterns.

CRL allows agents to be defined across multiple environments represented by different MDPs—a challenge for traditional RL definitions.
For instance, permanent road construction alters content dynamics in traffic monitoring, impacting the transition probability distribution and changing the MDP.
If the MDP shifts during or after training, traditional RL agents, such as those offline-trained in existing methods ~\cite{fang2017qos, coviello2021magicpipe, she2024earlyexit, zhang2024bcedge}, cannot adapt to the new environment.

Based on the notation on Abel et al.~\cite{abel2024continual} the state space $S$ is called observations $O$, which combined with actions $A$ create histories $h\in H$ of sequential pairs $h=o_0a_0...o_na_n$ for $o_n\in O$ and $a_n\in A$.
An agent is represented as a function $\lambda \in \Lambda: H \rightarrow \Delta(A)$, with $\Delta$ representing a probability distribution over a countable set.
The environment is a function $e \in E: H \times A\rightarrow \Delta(O)$ that can capture the MDP (\autoref{subsec:method_mdp}).
The learning process is described as agents searching for the set of optimal agents $\Lambda^*\subset\Lambda$, and the notation $\Lambda_1 \vdash_e \Lambda_2$ refers to $\Lambda_1\subset\Lambda$ converging to $\Lambda_2\subset\Lambda$ by observing $e$. 
Proof that the presented \iagent{} is a valid CRL problem is presented in the Supplementary Material.

\textbf{Extension to FCRL.}
The formulation of CRL can be extended to FCRL by defining it over multiple instances.
Within every round of federated learning, the model sent from server to instance $i$ should be interpreted as the basis $\Lambda_B$ for the local CRL problem $(e_i, v_i, \Lambda, \Lambda_B)$.
The aggregated model parameters serve as a basis that is not optimal without personalization and quickly adapted with continual learning.

This notation captures how every \iagent{} solves it's personal CRL problem based on shared general knowledge.
While an instance can encounter different environments and performance evaluations, the set of all agents $\Lambda$ is identical.
In \systemname{} the performance function is also identical ($\forall i_1,i_2: v_{i_1}=v_{i_2}$), forming a special case of FCRL problem where instances target an identical goal.

\begin{figure}[t]
    \centering
    \includegraphics[width=\linewidth]{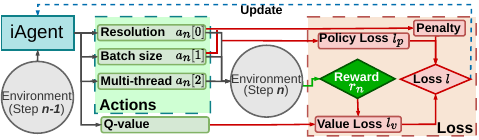}
    \caption{The training framework of \systemname{}'s \iagent{}(\autoref{fig:iagent-arch}). }
    \label{fig:model_architecture}
    % \vspace{-1.5em}
\end{figure}

\textbf{Loss Architecture.}
The loss calculated based on the steps in the current episode, using the $ratio$ of explored to exploited actions.
Besides policy and value loss, \iagent{} receives a direct penalty for actions 1 (resolution) and 3 (multi-threading), which is visualized in \autoref{fig:model_architecture}. 
The total loss $l$ is determined as:
\begin{equation}
\label{eq:loss}
  l = l_{p} + l_{v} + \omega~\cdot~ \frac{1}{n} \sum_n \left(a_n[0] + a_n[2]\right)
\end{equation}
where $\omega$ is a hyperparameter to adjust the penalty. 
The policy loss $l_p$ is the average of the clipped ratio $\epsilon \cdot ratio$, balancing exploration, multiplied with the Generalized Advantage Estimation to learn transition probability distribution~\cite{schulman2017proximal}.
The exponential of the negative reward is included as a factor to provide more direct feedback of the total reward value and fast continual adaptation to slight changes.
\begin{equation}
\label{eq:policy_loss}
  l_{p} = \frac{1}{n} \sum_{n} \min\big(\epsilon \cdot \text{ratio}, \text{ratio}\big) \cdot \big(\text{GAE} + e^{-r_n}\big)
\end{equation}
The mean squared error $mse()$ between the estimated Q-values and the collected rewards forms the value loss $l_v$.
By optimizing this, \iagent{} learns policies with better performance.
This estimation is also used for sharing knowledge of the overall environment trends across federated agents.
\begin{equation}
  l_{v} = \mathrm{mse}(Q(s,a)_n,r_n)
\end{equation}

\par
The reward function is complex, and adding more components makes it harder to optimize and balance the parameters effectively. 
To address this, the loss penalty serves as a way to ensure the batch size is optimized first, while other actions are only used when the improvement in the primary objectives is substantial enough to justify their trade-offs.
The secondary goals of accuracy and resource consumption are more challenging to evaluate from a local perspective, without labels to assess accuracy and resource consumption that may seem acceptable to one task but is detrimental to others.

\begin{algorithm}[t]
    \SetKwFunction{Main}{Main}
    \SetKwProg{Fn}{Function}{:}{}
    \SetAlFnt{\footnotesize} % Reduced font size
    \footnotesize % Apply smaller font size to the entire environment
    \Fn{\Main{}}{
        \footnotesize
        \textbf{select} and \textbf{await} selected clients \\
        \For{$l \in \text{base\_network layers}$} {
            agg\_layers.add($l$) \\
        }
        \For{$m \in $ \text{client models} $M$} {
            \For{$l \in \{layer_1, layer_2, layer_{\text{value}}\}$} {
                agg\_layers[$l$] $\mathrel{+}= m[l]$ \\
            }
            \For{$l \in \{layer_{a[0]}, layer_{a[1]}, layer_{a[2]}\}$} {
                $factor = \left(\textsc{loss}_l - \frac{\textsc{loss\_total}}{|M|}\right)^{-1}$ \\
                agg\_layers[$l$] $\mathrel{+}= factor \cdot m[l]$ \\
                $\textsc{loss\_total} \mathrel{+}= \textsc{loss}_l$ \\
            }
        }
        agg\_layers $\mathrel{/}= |M| + 1$\\
        \For{$m \in $ \text{client models} $M$} {
            \For{$l \in \{layer_1, layer_2, layer_{\text{value}}\}$} {
                $m[l].\text{load}(agg\_layers[l])$\\
            }
            send $m$ to client \\
        }
        base\_network.load(agg\_layers)
    }
    \caption{Agent-specific Aggregation - Server}
    \label{algo:federated_aggregation_server}
\end{algorithm}

\textbf{Overhead Minimization.}
Deploying CRL on embedded devices introduces overhead that can reduce the performance of the VA system.
Training neural networks is particularly challenging due to limited resource availability.
To address this, \iagent{}s employ a loss gate that executes back-propagation only when the improvement is significant. If the loss magnitude falls below a specified threshold, the network update is minimal and can be skipped.
However, to prevent the learning process from stagnating in sub-optimal positions, the FL update is always executed, as described in the next section.

Between RL updates, storing \textit{experiences} increases memory consumption—a critical issue on embedded devices.
Each agent maintains its experiences in a fixed-sized buffer, which imposes an upper limit on memory usage.
This issue is even more pronounced in RL systems performing offline updates over extended periods, as they require large buffers to collect numerous experiences.
For example, for each update, BCEdge \cite{zhang2024bcedge} and DDQN \cite{she2024earlyexit} store over 5000 experiences.

In \systemname{}, online CRL training allows the buffer to remain small and to be emptied frequently, significantly reducing memory overhead.
Additionally, \iagent{}’s buffer is populated based on experience diversity to maximize training efficiency.
After each forward pass, diversity $d$ is calculated as follows:
\begin{equation}
\label{eq:diversity}
d = \alpha \cdot D_M(s_n, s_{n-1}, \cdots, s_0) + \beta \cdot D_{KL}(\pi)
\end{equation}
where $D_M$ is the Mahalanobis distance between the new state and stored states, which emphasizes novelty, and $D_{KL}$ is the KL-Divergence of policy distributions, which captures deviations in action spaces.
This lightweight buffer design eliminates sequential dependencies between experiences and improves their IID distribution.
Thus, \iagent{}s do not require an additional replay buffer of past episodes like BCEdge~\cite{zhang2024bcedge}.

\subsection{Federated Reinforcement Learning (FRL)}
\label{subsec:method_federated_optimization}
\textbf{Knowledge Aggregation.}
As shown in \autoref{algo:federated_aggregation_server}, each federated update step start by selecting clients for aggregation.
Next, local weights and experiences are collected after a last local update (line 2).
During aggregation, only the backbones and Q-value head are aggregated equally (lines 6-7, 12).
These layers are designed to extract features from a state and provide a general understanding of the environment that remains consistent across agents.
Equal aggregation ensures that all agents contribute equally to this shared knowledge, preventing any single agent from dominating the model’s understanding.
A weighted aggregation, in contrast, could prioritize local information from specific agents, leading to imbalances and potentially diminishing the performance of other agents.

\par
On the other hand, action heads are aggregated using a loss-based aggregation across all agents with the same output dimensions (lines 8-11).
It is essential to store different action heads based on their dimensions since weights for selecting batch sizes of 1-64 or 1-16 are not directly comparable.

\begin{algorithm}[t]
    \SetKwFunction{Main}{Main}
    \SetKwProg{Fn}{Function}{:}{}
    \SetAlFnt{\footnotesize} % Reduced font size
    \footnotesize % Apply smaller font size to the entire environment
    \Fn{\Main{}}{
        \footnotesize
        $m_{local}$ = \textit{local\_update}$(m_{local})$; 
        send $m_{local}$ to server\\
        \While{\textbf{await} $(m_{aggregated})$ from server}{
            continue inference\\
            \textit{history\_states} += \textit{state}; \textit{history\_actions} += \textit{action}
        }
        $\textsc{policy} = m(\textit{history\_states})$\\
        $\textsc{loss} = \text{neg\_log\_likelihood}(\textsc{policy}, \textit{history\_actions})$\\
        $m_{aggregated}$.freeze$(layer_1, layer_2, layer_{value})$\\
        $m_{aggregated}$.update$(\textsc{loss})$; $\iagent{}$.load$(m_{aggregated})$
   }
    \caption{Agent-specific Aggregation - Client}
    \label{algo:federated_aggregation_client}
\end{algorithm}

\par
After this, the updated network is transferred back to the agents, replacing the current network (line 17).
During this process, the agents do not perform any local updates, and any experiences collected during this time are discarded.
The latest aggregated network, which includes the updated action heads, is then stored on the server (line 18). 

\par
\textbf{Action Head Fine-Tuning.}
Using the aggregated layers directly to make decisions can lead to unpredictable performance because the output from the backbone is not aligned with the action heads.
To address this, we fine-tune the model with experiences collected locally at each agent, focusing solely on the policy loss while freezing the value head and backbone (\autoref{algo:federated_aggregation_client} lines 6-9).
This fine-tuning step, is performed on the edge devices, as it is faster than a regular update, and therefore should not introduce additional overhead.
Also fine-tuning at the server would require sending local experiences to the server, adding additional network overhead.

\par
\textbf{Large-Scale FL.}
\systemname{} combines hierarchical FL with client selection to minimize network overhead caused by the frequent exchange of model parameters. 
The Edge inherently represents a hierarchical network structure where edge devices are co-located with cameras and connect to a local cluster edge server.
This edge server can then communicate with other clusters or the cloud to distribute workload.

\par
This topology inherently forms clusters of devices, which can be leveraged to reduce the number of participating devices. 
Additionally, client selection that considers memory, computing availability, and data heterogeneity as FedHybrid~\cite{tam2024fedhybrid} reduces the risk of stragglers. 
The considerations of FedHybrid align well with \systemname{}s efforts to reduce device overhead and prioritizing diversity.
To further lower network overhead, we extend the utility function proposed in \cite{tam2024fedhybrid} by incorporating bandwidth (normalized to $10 Mbit/s$) for client $c$ as follows:
\begin{equation}
TotalUtil(c) = Util(c)_\text{\cite{tam2024fedhybrid}} * \sqrt{Bandwidth(c) / 10}
\end{equation}

The \textit{System Controller} will determine the participation of a device, considering the utility sum of all agents in a single device based on $TotalUtil(c)$. 
A chosen \textit{Device Control} makes the final decision, which of the local \iagent{}s will participate in the next FL round based on memory availability.

\par
Taking these factors into account, \systemname{} carries out fine-grained client selection within each cluster.
Once multiple aggregation rounds are completed within a cluster, the updates are shared with other clusters through the cloud, as proposed in~\cite{liu2020client}.
This step follows the same aggregation process used at the edge server and the same number of local rounds.

\begin{figure}[t]
    \centering
    \includegraphics[width=\linewidth,trim=15 10 15 8, clip]{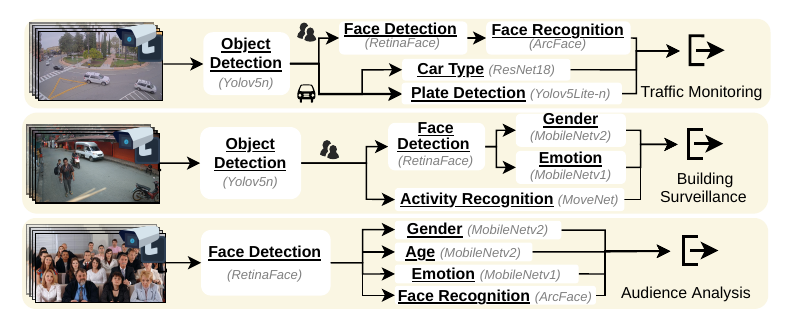}
    \caption{Sample pipelines for traffic, surveillance, and audience analysis.}
    \label{fig:pipelines}
    % \vspace{-1em}
\end{figure}

% \begin{table}[t]
% \vspace{0.1em}
% \centering
% \caption{Summary of \systemname{}'s implementation and training parameters.}
% \vspace{-0.5em}
%     \label{tab:fcpo-parameters}
%     \begin{tabular}{@{\hspace{1pt}}m{1.5cm}|@{\hspace{3pt}}m{6.5cm}@{}}
%     \hline
%     \textbf{Parameters}& \textbf{Description} \\\hline
%     $n_s = 10$ & Number of steps in each episode \\\hline
%     $LR = 10^{-3}$ & Learning rate of \iagent{} \\\hline
%     $\vartheta=1.1$ $\ \ \ \ $ $\varsigma=10$ $\ \ \ \ $ $\varphi = 2$ & Weights to calculate the reward (\autoref{eq:reward-function}) \\\hline
%     $\gamma,\lambda = 0.1$ & Weights of reward temporal relation (\autoref{eq:q-function} + GAE) \\\hline
%     $\omega = 0.2$& Weight of loss penalty (\autoref{eq:loss}) \\ \hline
%     $\epsilon = 0.9$ & Clip value in policy loss (\autoref{eq:policy_loss}) \\\hline
%     $\alpha,\beta=0.5$ & Weights to calculate experience diversity (\autoref{eq:diversity}) \\\hline
%     \end{tabular}
%     \vspace{-1.5em}
% \end{table}

\section{Evaluation}

%\begin{figure}
%    \centering
%    \includegraphics[width=\linewidth]{figures/test_bed.jpg}
%    \caption{\systemname{}'s testbed devices}
%    \label{fig:testbed}
%\end{figure}
\label{sec:evaluation}
In our evaluation, we assess \systemname{} on a real-world testbed and aim to answer the following questions:
% (comprising a server equipped with four consumer-grade NVIDIA GPUs RTX 3090 and 12 heterogeneous devices (3x Jetson Xavier AGX, 5x Jetson Xavier NXs, 3x Jetson Orin Nanos, and 1x "On-Premise" PC equipped with a GTX 1080TI).
\begin{enumerate}[label=\textbf{\textbullet \ Q\arabic*}, leftmargin=9mm]
    \item (\ref{subsubsec:eval_performance}): \textit{Does \systemname{} outperform other Edge VA systems with and without local optimization?}
    \item (\ref{subsubsec:eval_fl}): \textit{How fast and reliable is \systemname{}s FL procedure?}
    \item (\ref{subsubsec:eval_tight_latency}): \textit{Can \systemname{} adapt to strict real-time SLOs?}
    \item (\ref{subsubsec:eval_warm_start}): \textit{Does \systemname{} enable measurable warm starts?}
    \item (\ref{subsubsec:eval_overhead}): \textit{How much overhead does \systemname{} incur on the devices? And does that prohibit scalability?}
    \item (\ref{subsubsec:eval_continual}): \textit{Can \systemname{} adapt quickly in drastically changing environments? And if so, how does CRL help?}
    \item (\ref{subsubsec:eval_convergence}): \textit{How does FRL benefit learning convergence?}
\end{enumerate}

\subsection{Experimental Methodology}
\subsubsection{\textbf{Real-world Testbed}}
We evaluate \systemname{} with an edge server with 4 consumer-grade GPU NVIDIA RTX 3090 and 12 heterogeneous devices consisting of 3 Jetson Xavier AGXs, 5 Jetson Xavier NXs, 3 Jetson Orin Nanos and an "On-Premise" desktop PC equipped with a GTX 1080Ti.
We emulate real-world network bandwidth conditions using the data transfer benchmark of an Irish 5G dataset ~\cite{raca2020beyond}.

\par
\subsubsection{\textbf{System Implementation}}
\systemname{}'s implementation is build based on PipelineScheduler~\cite{pipelinescheduler} in C++ and adds over 5,000 new lines of code to the system.
We deploy each inference model within a container and leverage a microservice architecture to harness its inherent flexibility and robustness, with Docker serving as the container runtime.
However, \systemname{} can also operate in a monolithic architecture with minimal modification.
We implement and run the \textit{Controller} as a separate process on the server to supervise the operation of the entire cluster.
Since the focus of \systemname{} is continual local adaptation, we utilize \cite{nguyen2025octopinf} to make global scheduling decisions at the \textit{Controller}, distributing the workload across devices every 5 minutes. 
On each edge device and the server where inference containers are hosted, we deploy a \textit{Device Agent} to manage and monitor the containers.

\begin{table}[t]
\vspace{0.1em}
\centering
\caption{Summary of \systemname{}'s implementation and training parameters.}
\vspace{-0.5em}
    \label{tab:fcpo-parameters}
    \begin{tabular}{@{\hspace{1pt}}m{2.3cm}|@{\hspace{3pt}}m{6.1cm}@{}}
    \hline
    \textbf{Parameters}& \textbf{Description} \\\hline
    $n_s = 10$ & Number of steps in each episode \\\hline
    $LR = 10^{-3}$ & Learning rate of \iagent{} \\\hline
    $\vartheta, \varsigma, \varphi=1.1, 10, 2$ & Weights to calculate the reward (\autoref{eq:reward-function}) \\\hline
    $\gamma,\lambda = 0.1$ & Weights of reward temporal relation (\autoref{eq:q-function} + GAE) \\\hline
    $\omega = 0.2$& Weight of loss penalty (\autoref{eq:loss}) \\ \hline
    $\epsilon = 0.9$ & Clip value in policy loss (\autoref{eq:policy_loss}) \\\hline
    $\alpha,\beta=0.5$ & Weights to calculate experience diversity (\autoref{eq:diversity}) \\\hline
    \end{tabular}
    % \vspace{-1.5em}
\end{table}

\par
Within each container, we deploy each inference model in a single process.
We use OpenCV (4.8.1) for image (video frame) pre- and post-processing tasks.
Different inference engines, such as TensorRT, ONNX, TF Lite, and OpenVINO, are supported in an easy plug-and-play manner.
For the experiments in this paper, we use TensorRT (8.4.3.1) to load and run inference models.
As described in \autoref{sec:fcpo}, we attach a lightweight \iagent{} (implemented with LibTorch) to each inference process to optimize its operation.
The parameters used for the experiments are detailed in \autoref{tab:fcpo-parameters}. The code base  can be applied to various visual inference tasks requiring little modification.

\subsubsection{\textbf{Edge VA Workloads}}
As shown in \autoref{fig:pipelines}, we use traffic monitoring, building surveillance, and audience analysis as three distinct representative edge VA applications to evaluate \systemname{}.
We set a strict end-to-end SLO of 250ms for all pipelines to reflect the prompt nature of Edge VA applications and services.
For data, we collected a total of 23 continual 4-hour videos from real-world public online streams to represent 23 data sources with diverse object distributions and content dynamics.
During the experiments, the videos are streamed at 15 FPS to simulate real-time video sources across devices.

\par
We also leverage the vehicle tracking data from the 2022 AI City Challenge~\cite{Naphade22AIC22} to evaluate \iagent{}, which was unsupervisedly trained on our collected dataset, in its ability to adapt to a new domain \autoref{fig:warm_start_eval}.
We use 9 6-min videos at 10 FPS.

\subsubsection{\textbf{Baselines}}
For all experiments, we leverage OctopInf~\cite{nguyen2025octopinf} to distribute the workload across multiple devices every 5 minutes.
Our goal is to demonstrate that combining \textit{infrequent global scheduling} with \textit{continual local optimization} provided by \systemname{} can significantly improve performance.

\begin{itemize}[wide] 
    \item \textit{BCEdge}~\cite{zhang2024bcedge} represents state-of-the-art RL-based throughput optimization.
    We selected BCEdge as a baseline because, to the best of our knowledge, it is the most recent work on this topic and exhibits the most desirable qualities among the SOTAs compared in \autoref{Tab:sota-comparison}.
    We deploy one agent per device, which was trained offline using profiling results, and is updated every 7000 experiences. To prevent catastrophic failures by excessive GPU resource requests, we limit the concurrency and shared memory actions to only two configurations each.
    \item \textit{OctopInf}~\cite{nguyen2025octopinf} serves as a global scheduling baseline without local optimizations, but global parameter settings.  
    \item \textit{Distream}~\cite{zeng2020distream} is a popular baseline without any detailed runtime optimization of parameters like batch size. 
\end{itemize}

\begin{figure}
    \centering
    \begin{minipage}[h]{.34\linewidth}
        \begin{subfigure}{\textwidth}
            \centering
            \includegraphics[width=\textwidth, trim=10 10 10 10, clip]{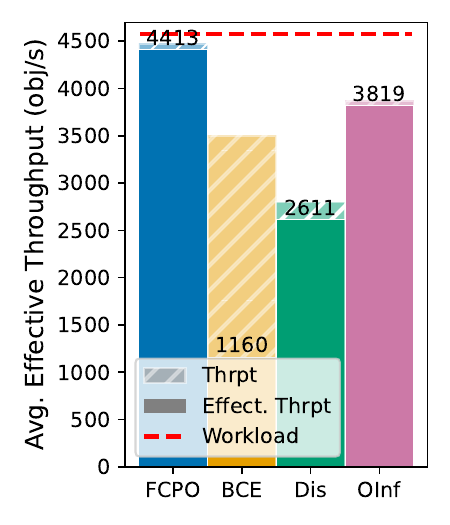}
            \vspace{-1.4em}
            \caption{Effective throughput}
        \end{subfigure}
        \begin{subfigure}{\textwidth}
            \centering
            \includegraphics[width=\textwidth, trim=0 0 0 0, clip]{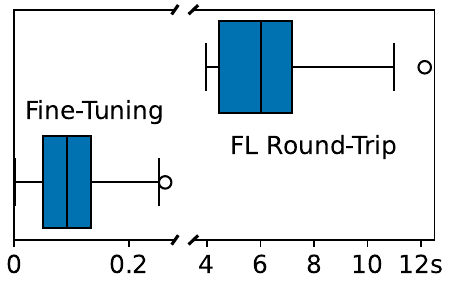}
            \vspace{-1.4em}
            \caption{FL roundtrip latency}
        \end{subfigure}
    \end{minipage}
    \begin{minipage}[h]{.64\linewidth}
        \centering
        \begin{subfigure}{\textwidth}
            \centering
            \includegraphics[width=\textwidth, trim=5 10 10 10, clip]{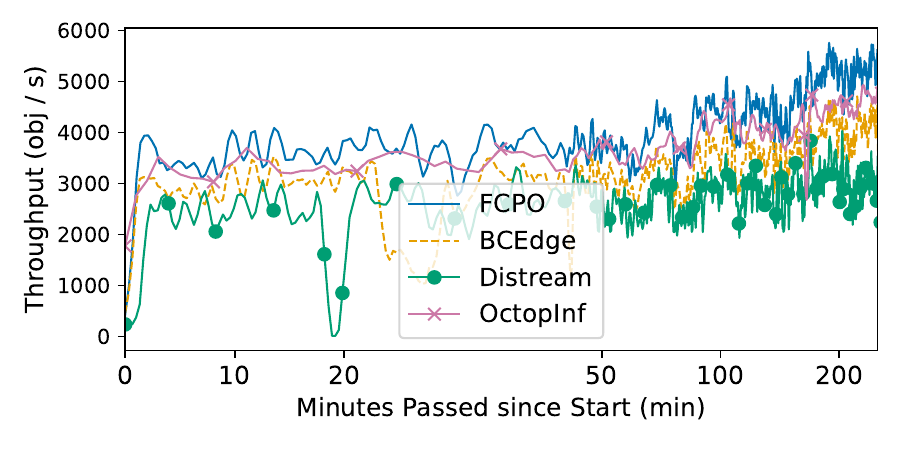}
            \vspace{-1.4em}
            \caption{Averaged total end-to-end throughput.}
        \end{subfigure}
        \begin{subfigure}{\textwidth}
            \centering
            \includegraphics[width=\textwidth, trim=5 10 10 0, clip]{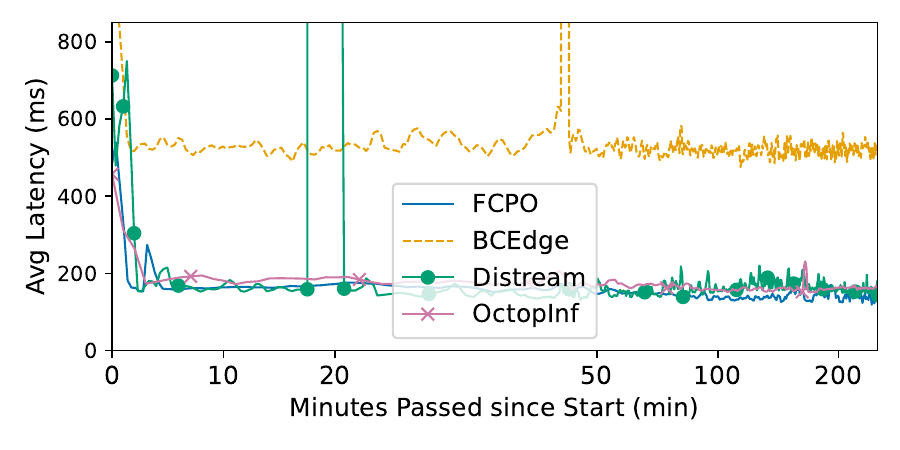}
            \vspace{-1.4em}
            \caption{Average end-to-end latency of arrived requests.}
        \end{subfigure}
    \end{minipage}
    \hfill
    \caption{End-to-end system performance comparisons over 4h.}
    \label{fig:main-results-throughput}
\end{figure}

\subsubsection{\textbf{Metrics}}
Every reported metric is the average value over three runs.
We validate \systemname{} by comparing its end-to-end performance against baselines using the following:

\begin{itemize} [wide]
    \item \textit{End-to-End Latency.} This is the time elapsed from the generation of a video frame at the source to the moment its inference results arrive at the designated \textit{sink}, including all network, queuing, and processing latencies.
    It measures the responsiveness of the system in handling video streams.

    \item \textit{Throughput.} Our scenarios involve pipelines that analyze attributes of objects-of-interest using inference models.
    Therefore, we define \textit{\textbf{throughput}} as the total number of objects that are analyzed by the pipelines within one second. 
    However, results that arrive late—where their end-to-end latency exceeds their SLO (set at 250ms)—are no longer useful. Thus, we also measure the \textit{\textbf{effective throughput}}, which is \textit{the total number of inference results that arrive on time within one second}.
    This number highlights the degree of real-time processing.
\end{itemize}

\par
To prove the scalability and real-time capability of \systemname{}, we analyze the overhead incurred by \iagent{} using the following metrics for RL and system resources:
\begin{itemize}
    \item \textit{Convergence Speed.} Convergence refers to the state of accurately approximating optimal decision. Its speed indicates how quickly \iagent{} learns the environment.
    \item \textit{Memory Consumption.} Memory allocated to \iagent{}.
    \item \textit{Power Consumption.} \iagent{}'s power consumption on edge devices.
    \item \textit{Decision Latency.} Time taken to choose a single action.  
    \item \textit{Training Latency.} This is the time required to update the network on-device locally without FL. 
\end{itemize}

\subsection{Results}

\subsubsection{\textbf{Processing Throughput and Inference Latency Improvements}}
\label{subsubsec:eval_performance}

\autoref{fig:main-results-throughput} shows the overall throughput performance of \systemname{}.
\systemname{} significantly outperforms the baselines in terms of \textbf{throughput} and \textbf{effective throughput}.
\systemname{} implements local continual optimization on top of OctopInf's global workload distribution.
By continually adapting to the environment via actions such as batch size adjustment \systemname{} manages to output nearly 2 times the throughput and better latency (\autoref{fig:main-results-throughput}d) most of the time, which proves the validity of our framework.
However, this is not the case for BCEdge, which also employs a local adaptation approach on top of OctopInf (\autoref{fig:main-results-throughput}a and c).
By adjusting batch sizes, compared to Distream, BCEdge reduces the number of queue drops and thus achieves higher throughput.
However, only a small portion of this is effective throughput because its average latency is significantly worse (\autoref{fig:main-results-throughput}d).
While increasing throughput, batched inference raises the latency of each inference request \cite{nigade2022jellyfish}. 
The effect is compounded over multiple stages of the pipeline, resulting in nearly 2.5 times higher latency.

\begin{figure}
    \centering
    \begin{subfigure}{.493\linewidth}
        \centering
        \includegraphics[width=\textwidth, trim=10 5 5 0, clip]{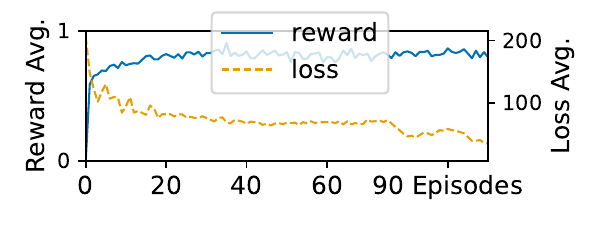}
        \vspace{-1.5em}
        \caption{\systemname{} (\textit{online training})}
    \end{subfigure} 
    \begin{subfigure}{.493\linewidth}
        \centering
        \includegraphics[width=\textwidth, trim=10 5 5 0, clip]{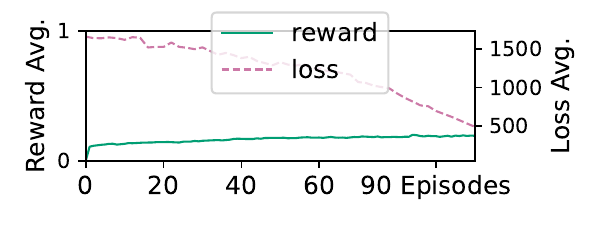}
        \vspace{-1.5em}
        \caption{BCEdge~\cite{zhang2024bcedge} (\textit{offline training})}
    \end{subfigure}
    \caption{Learning performance with averaged loss and rewards.}
    \label{fig:main-results-reward-loss}
    \vspace{-0.5em}
\end{figure}

\par
We further shed light on the root cause by analyzing \autoref{fig:main-results-reward-loss}.
The reward and loss of \systemname{} show gradual improvement and importantly noticeable fluctuation as it continually tries to adapt to the ever-changing environment dynamics.
Contrary, BCEdge is \textbf{trained offline} with profiling data and its reward quickly converges because profiling data is obviously less diverse in workload patterns and cannot capture all the conditions of devices as well as the interactions among various applications (processes) running on the devices.

\par
Another cause is that there is only one BCEdge agent on each device to make decisions for all workloads, which becomes the bottleneck.
Contrary, \systemname{} has one light-weight \iagent{} for each workload increasing the timeliness of its decision.
We will further discuss the scalability shortly.

\subsubsection{\textbf{Federated Learning Latency}}
\label{subsubsec:eval_fl}
The FL latencies in \autoref{fig:main-results-throughput}b show that the average FL round completes in approximately 4–8 seconds.
This aligns with findings from a recent study~\cite{hayek2025federated}, where an FL round using FedAvg~\cite{mcmahan2017communication} over 5G networks took 43 seconds with 6 nodes and a model size of approximately 3MB.
In contrast, the \iagent{} model is only 53KB in size, though it is aggregated across more nodes.
Despite this, the FL optimization in \systemname{} achieves efficient round-trip times—from sending weights to receiving the aggregated model.
Because local \iagent{}s continue operating during this period, FL latency does not hinder real-time processing.
Furthermore, on-device fine-tuning after aggregation introduces negligible overhead—less than 300ms across all devices—which is still below the highest update latency shown in \autoref{fig:fcpo_overhead_eval}e.
Thus, \iagent{} can reliably support continual adaptation in real-world deployments without compromising responsiveness.

%\subsubsection{Performance in Real Network Conditions}
%\label{subsubsec:eval_network}

%\begin{figure}
%    \centering
%    \includegraphics[width=0.9\linewidth, trim=10 10 10 0, clip]{figures/network-throughput.pdf}
%    \caption{Performance under real-world fluctuating 5G network bandwidths.}
%    \vspace{-0.5em}
%    \label{fig:restricted_network_eval}
%    \vspace{-0.5em}
%\end{figure}

%We emulate real-world network bandwidth conditions using the data transfer benchmark of an Irish 5G dataset ~\cite{raca2020beyond}.
%The results in \autoref{fig:restricted_network_eval} show how the effective throughput is maintained through different network conditions, only fluctuating slightly unless when encountering worst conditions.
%While \systemname{} itself does not include knowledge about the network conditions, by adjusting to the changed arrival rates at the server and optimizing the latencies it can still improve the effective throughput.
%Compared to BCEdge, \iagent{}s are able to improve much faster after an increase in network bandwidth, without noticeable delay.
%The occasional transmission of FL weights (\numprint{53}{KB}) is not latency-critical and does not show a measurable impact on the transmission of pipeline data.

\begin{figure}
    \centering
    \includegraphics[width=\linewidth, trim=10 10 10 0, clip]{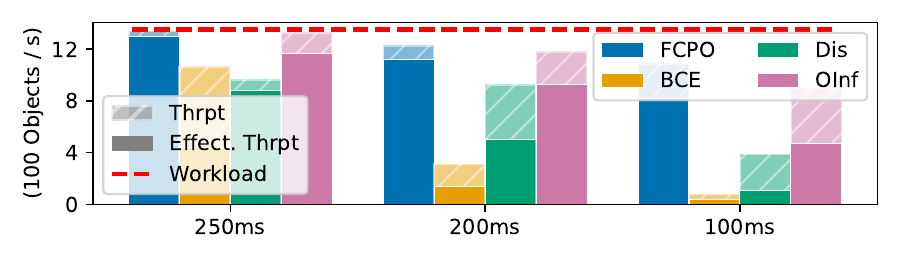}
    \caption{Throughput comparisons under increasingly restricted SLOs.}
    \vspace{-0.5em}
    \label{fig:strict_realtime}
    \vspace{0.75em}
\end{figure}

\subsubsection{\textbf{Adaptation to Stricter Real-Time Requirements}}
\label{subsubsec:eval_tight_latency}
We evaluate \iagent{}'s ability to adapt to real-time constraints by tightening the SLO from 250ms to 200ms and 100ms, respectively.
As shown in \autoref{fig:strict_realtime}, both Distream and OctopInf experience a severe drop in performance.
This is because they make periodic scheduling decisions based solely on average workload statistics from the previous period.
Under stricter SLOs, the system dynamics become significantly more volatile, requiring immediate and responsive adaptation.

By incorporating the end-to-end SLO as a state variable, \iagent{} learns to associate tighter latency requirements with appropriate system configurations, enabling it to respond more effectively to increased urgency and maintain better performance.
In contrast, although BCEdge incorporates SLO information into its reward function, it fails to capture the diversity of environment patterns, often leading to suboptimal decisions.
As a result, it accumulates only minor rewards (\autoref{fig:main-results-reward-loss}), which are insufficient to drive effective learning.

\subsubsection{\textbf{Analysis of Warm Starts}}
\label{subsubsec:eval_warm_start}

\begin{figure}[t]
    \centering
    \includegraphics[width=\linewidth, trim=10 10 10 0, clip]{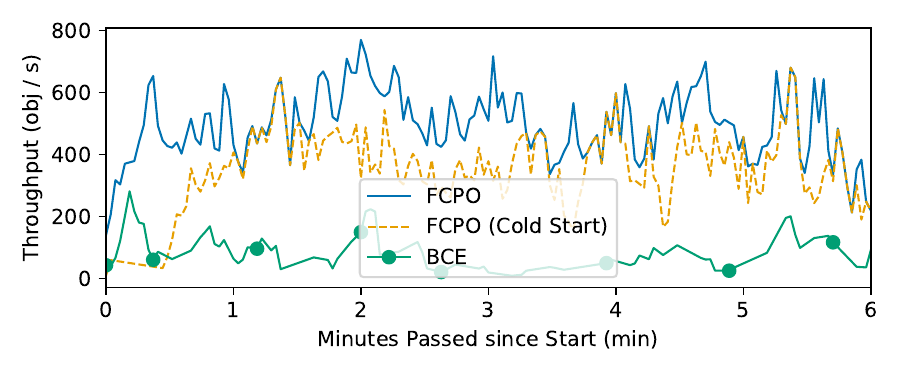}
    \caption{Performance on a traffic dataset \cite{Naphade22AIC22} with \textit{out-of-distribution} workload patterns.}
    \vspace{-1em}
    \label{fig:warm_start_eval}
    \vspace{1em}
\end{figure}

In real-world scenarios, environmental changes can occur both gradually and abruptly.
To evaluate \systemname{}'s ability to achieve a "warm start" even during abrupt changes, we replace the data sources with videos from the AI City Challenge dataset~\cite{Naphade22AIC22}, which exhibits \textit{out-of-distribution} workload patterns.
As shown in \autoref{fig:warm_start_eval}, the trained \iagent{} maintains high throughput throughout the experiment.
We then increase the difficulty by evaluating a blank \iagent{} (cold start).
Although it initially exhibits low throughput, it quickly adapts to the new environment and, by the end of the experiment, achieves performance only slightly below that of the warm-start \iagent{}.
In contrast, BCEdge performs worse than it did on the original dataset and fails to adapt to the new environment.
These results demonstrate that \iagent{} not only adapts effectively to previously unseen conditions but also supports both warm and cold start scenarios, making it highly suitable for deployment in dynamic, real-world settings.

\subsubsection{\textbf{Analysis of On-Device Agent Overhead}}
\label{subsubsec:eval_overhead}

\par
\begin{figure}
    \centering
    \begin{subfigure}{.32\linewidth}
        \centering
        \includegraphics[width=\textwidth, trim=0 5 0 10, clip]{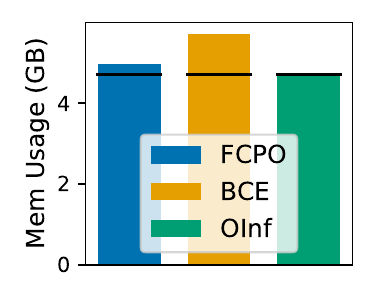}
        \vspace{-1.5em}
        \caption{Average \textit{device} memory consumption}
    \end{subfigure}
    \begin{subfigure}{.32\linewidth}
        \centering
        \includegraphics[width=\textwidth, trim=0 5 0 10, clip]{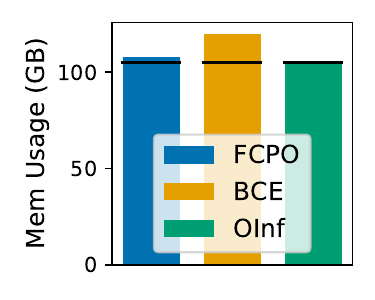}
        \vspace{-1.5em}
        \caption{Average \textit{server} memory consumption}
    \end{subfigure}
    \begin{subfigure}{0.32\linewidth}
        \centering
        \includegraphics[width=\textwidth, trim=0 5 0 0, clip]{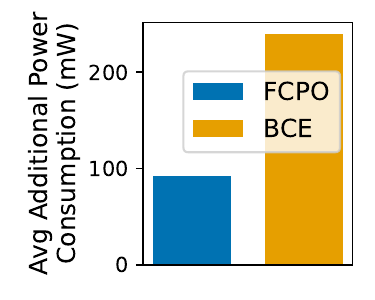}
        \vspace{-1.5em}
        \caption{Avg consumption for \textit{real-time local optimization}}
    \end{subfigure}
    \begin{subfigure}{.493\linewidth}
        \centering
        \includegraphics[width=\textwidth, trim=0 10 0 0, clip]{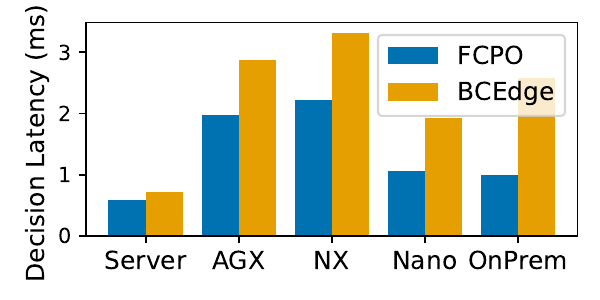}
        \vspace{-1.5em}
        \caption{Decision latencies per step.}
    \end{subfigure}
    \begin{subfigure}{.493\linewidth}
        \centering
        \includegraphics[width=\textwidth, trim=0 10 0 0, clip]{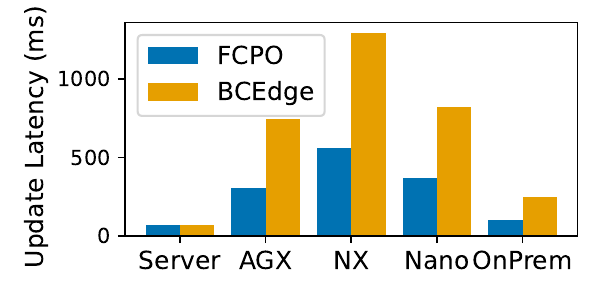}
        \vspace{-1.5em}
        \caption{Update latencies per episode.}
    \end{subfigure}
    \vspace{-1.5em}
    \caption{Agent overhead comparisons averaged over 4h. (\textit{Portions above the black line in (a) and (b) are memory consumption of RL agents})}
    \label{fig:fcpo_overhead_eval}
    % \vspace{-1em}
\end{figure}

To prove the scalability of \systemname{} show the overhead comparisons in \autoref{fig:fcpo_overhead_eval}.
Regarding memory consumption, all FCPO's \iagent{}s combined take less than 3\% of the total memory allocated for the experiments at both the Edge devices and the server.
This is owing to its lightweight structure and small fixed-size experience buffer.
On the other hand, even though there is only one BCEdge agent per device, it takes significantly more memory (up to $10\times$) due to a bulky structure and a large experience buffer (7000 experiences).
The BCEdge agent is deeper and wider compared to \iagent{}.
It also requires an additional branch to analyze the state value, which results in more intermediate layers, which in turn increases memory consumption.
Moreover, since BCEdge only needs to train offline the server, we disable it at the Edge devices to save resources for pipeline processing.
Otherwise, the memory consumption at the edge is expected to be even higher.

\par
We collect edge power consumption metrics through Jtop and calculate the agent overhead by subtracting the workload power consumption.
On average, BCEdge consumes more than twice as much energy as \systemname{} on each edge device.
This highlights how much more expensive an inference of BCEdge is on limited resources and how efficient \iagent{} operates.

\par
Regarding latencies, on each of the edge devices, BCEdge takes $1.5-2\times$ more time to make a decision due to its more complex model.
The training time of \systemname{} can safely execute until the next decision time after one second even on low-end devices like Orin Nano.
This result shows, how the light-weight design of \systemname{} is capable of real-time workload configuration.
The update latencies of BCEdge are collected offline, as it is not designed for online learning.
Running it`s training at the same time of inference will result in worse performance due to co-location interference~\cite{nguyen2025octopinf}.

\subsection{Ablation Study}

\begin{figure}
    \centering
    \begin{minipage}[c]{.696\linewidth}
        \begin{subfigure}{\textwidth}
            \centering
            \includegraphics[width=\textwidth, trim=10 10 10 0, clip]{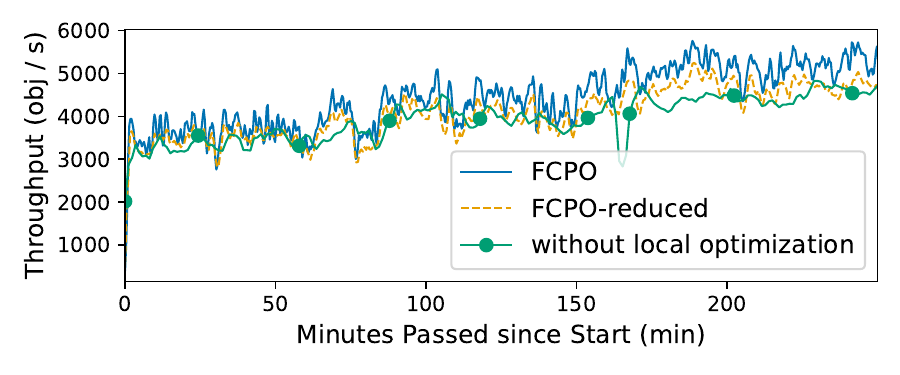}
        \end{subfigure}
        \begin{subfigure}{\textwidth}
            \centering
            \includegraphics[width=\textwidth, trim=10 10 10 10, clip]{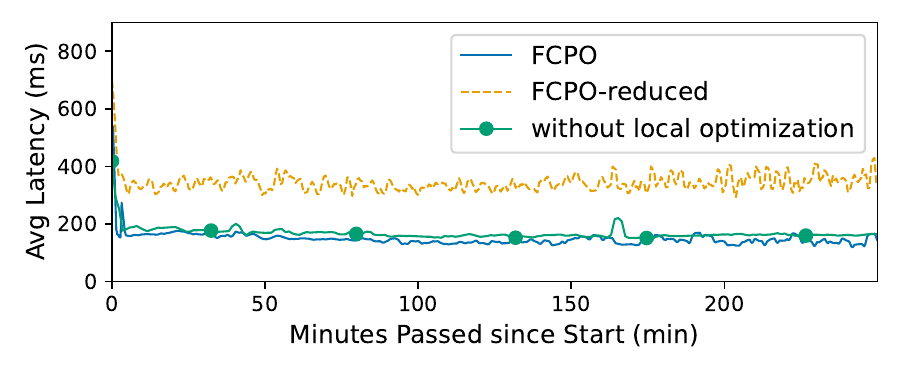}
        \end{subfigure}
    \end{minipage}
    \begin{minipage}[t]{.29\linewidth}
        \begin{subfigure}{\textwidth}
            \centering
            \includegraphics[width=\textwidth, trim=2 5 2 0, clip]{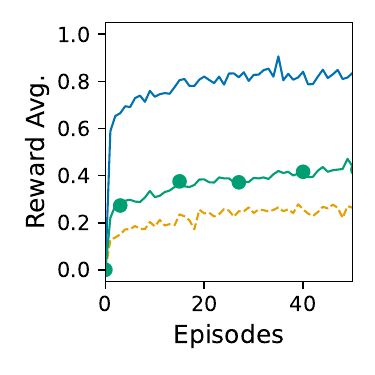}
        \end{subfigure}
        \begin{subfigure}{\textwidth}
            \centering            \includegraphics[width=\textwidth, trim=2 10 2 10, clip]{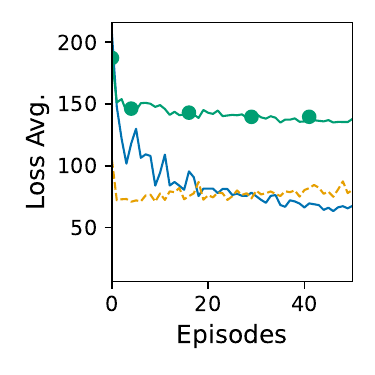}
        \end{subfigure}
    \end{minipage}
    \vspace{-0.25em}
    \caption{Ablation study comparisons with a reduced configurations.}
    \label{fig:ablation_study}
    % \vspace{-1.25em}
\end{figure}

To provide further insights into FCPO, we conduct an ablation study, where we remove two heads and use a single head for all three actions in the FCPO-reduced version.
As shown in \autoref{fig:ablation_study}, removing certain aspects of \systemname{} leads to significantly worse performance.
Specifically, deploying \iagent{} at the server making optimizations every 5 minutes results in less responsive updates, leading to suboptimal decisions. 
On the other hand, \systemname{}-reduced, while still able to perform updates, struggles to understand the action space, resulting in low rewards and high latency.
These findings support our argument that a single head introduces a highly complex action space, where \iagent{} fails to effectively explore its decisions.

\subsubsection{\textbf{Benefits of Continual Learning on Performance}}
\label{subsubsec:eval_continual}

\begin{figure}[t]
    \centering
    \includegraphics[width=\linewidth, trim=0 10 0 10, clip]{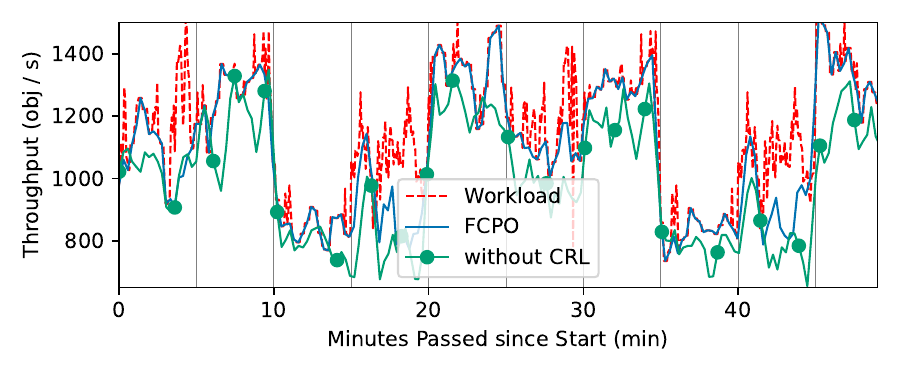}
    \caption{Impact of Continual Learning on Effective Throughput. (\textit{Vertical lines indicate a significant context switch}).}
    \label{fig:continual_evaluation}
    % \vspace{-1.7em}
\end{figure}

\begin{figure}[t]
    \centering
    \begin{subfigure}{.493\linewidth}
        \centering
        \includegraphics[width=\textwidth, trim=10 10 10 10, clip]{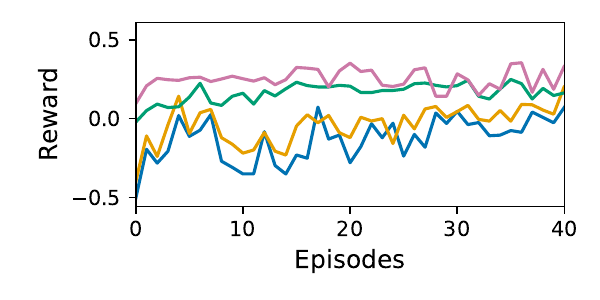}
    \end{subfigure}
    \begin{subfigure}{.493\linewidth}
        \centering
        \includegraphics[width=\textwidth, trim=10 10 10 10, clip]{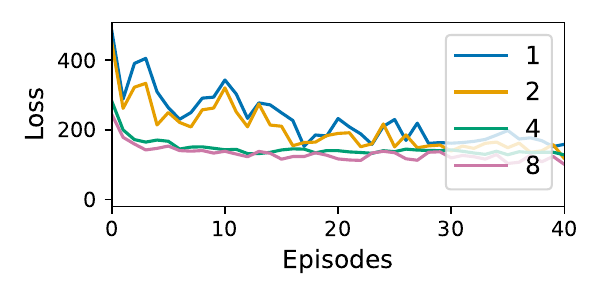}
    \end{subfigure}
    \vspace{-1.5em}
    \caption{Average loss and reward with different FRL instance counts.}
    \label{fig:federated_convergence}
    \vspace{-0.5em}
\end{figure}

To capture the performance impact of including CRL in the design, we concatenate multiple 5-minute video segments from different sources, causing the underlying distributions to change drastically.
We evaluate two \textit{trained} instances of \systemname{} at the same checkpoint over 50 minutes of traffic monitoring at 3 sources.
One agent is \textit{frozen} without CRL, while the other is allowed to \textit{learn} continually using CRL.
\autoref{fig:continual_evaluation} shows that, although the frozen agent is unable to adapt, it still makes near-optimal decisions because it was trained beforehand.
However, it is outperformed by the learning agent throughout the entire experiment.
The CRL agent shows better adaptation to workload spikes, but unfamiliar scenes like 40 to 45 minutes can still present a challenge.
While both agents follow similar patterns and have comparable performances at each context switch, the CRL agent can quickly adapt to the new context and perform significantly better until the next switch.
This proves the effectiveness of CRL in adapting to changing environments.

\subsubsection{\textbf{Convergence Speed with Federated Learning}}
\label{subsubsec:eval_convergence}

\autoref{fig:federated_convergence} illustrates the convergence speed, at varying numbers of pipelines used during FRL training in the traffic pipeline.
Results are averaged across all running models and all parameters are identical across configurations.
In the single-pipeline scenario (1) federated aggregation is disabled, the other configurations aggregate every second episode.

\par
As the number of pipelines—and consequently \iagent{}s—increases, both reward and the rate of loss reduction improve, with the most significant gains observed between 4 and 8 pipelines.
This suggests that learning scales effectively with more federated agents, though diminishing returns emerge at higher numbers.
This effect may reflect an upper limit on achievable improvements, as loss and reward converge quickly. 
Each additional pipeline introduces slight variations in dynamics, increasing the complexity of optimizing the global model. 
These variations are also the cause of learning divergence.
\textit{Instead of diverging}, FRL combines these differences into a more general scene understanding and smooth the learning curves, resulting in less performance fluctuations.

\section{Discussion}
\label{sec:discussion}

\subsection{Feasibility and Scalability in Real-world Environments}

In this section, we further examine the feasibility and scalability of \systemname{} in real-world environments, where sensitivity to system overhead, large-scale collaborative learning, and heterogeneity in applications and devices are key challenges.

\textbf{Overheads.}
From the experimental results, each device, on average, incurs less than $3\%$ total memory usage, less than $1.7\%$ total power consumption, and approximately $53$KB of network overhead every 300s for continual adaptation—resulting in more than a 13\% improvement in throughput.
These overheads are negligible, thanks to our lightweight design of \iagent{} and efficient Agent-specific Federated Learning (FL) aggregation mechanism.
This minimal resource footprint is crucial for practical real-world deployments, where computational and energy budgets are constrained.

% At first glance, our design, which assigns one \iagent{} per model, might appear non-scalable due to the added overhead of each instance.
% However, our memory analysis shows that the additional resource consumption is negligible ($<4\%$ total consumption).
% Both at the edge, where only a few models are deployed on average, and at the server, which handles a large number of models, the overhead is significantly smaller compared to deploying a device-wide agent like SOTAs 
% A key factor in this observation is the experience buffer and the lightweight design of \systemname{}.
% This minimal overhead is outweighed by the benefits of having a highly specialized local \iagent{}.
% \textbf{Future work.}

\textbf{Large-scale Collaboration. }
In dense Edge networks, vanilla Federated Learning (FL)—where all agents participate in every round—is typically infeasible due to the significant network and computation overhead imposed on the centralized server, which becomes a bottleneck.
To address this, \systemname{} is designed to selectively involve only a subset of agents with the most valuable experiences in each round, thereby maintaining constant system complexity.
Meanwhile, the remaining agents continue to perform local optimizations independently.
% Cooperation among agents may improve performance but also increase communication overhead and action selection latency.
% We believe the benefits of direct communication will diminish compared to the added overhead as the number of agents scales.
% Instead, we plan to integrate fast local optimization with broader global resource allocation.
% By combining device-level optimizations with system-wide coordination, we aim to balance performance and resource utilization.
% This hybrid approach is expected to maximize efficiency and scalability, enabling adaptive responses to dynamic workloads and device capabilities in real time. 
% One way to improve global end-to-end throughput through local optimization is global reward learning, which reduces development time by allowing the reward function to self-optimize and potentially outperform human-tuned parameters.

% \textbf{Future work.} We aim to enable even greater scalability through decentralized or hybrid Federated Learning (FL), where the aggregation of \iagent{}s could occur within local clusters or even directly on individual devices.
% While this concept has been explored in prior work, it has not been considered for real-time inference.
% Additionally, further investigation is needed to fully understand the trade-offs and overheads associated with such decentralized aggregation strategies.

\textbf{Application Heterogeneity. }
The state and action spaces of \systemname{} are tailored for VA applications.
However, its adaptation mechanism—Federated Continual Reinforcement Learning—is general and applicable to a broad range of tasks.
\systemname{} learns the environment state and selects actions such as batching, multi-threading, and resolution adjustment to maximize throughput.
Similar strategies are widely used in machine learning, including for time-series prediction and natural language processing.
For example, batching has proven effective for RNNs~\cite{gao2020batchrnn, holmes2019batchrnn} and Transformers~\cite{noauthor2025batchtransformer, noauthor2023batchllm}, both of which benefit from parallel processing.
This suggests \systemname{} can be extended to support such applications efficiently.

\textbf{Hardware and Software Heterogeneity. }
A key challenge to real-world scalability lies in the heterogeneous mix of hardware and software platforms from different manufacturers.
\systemname{} is built around generalizable actions whose effectiveness can be consistently evaluated across major platforms.
While batched inference is a relatively recent innovation from the deep learning era—unlike established techniques like resolution adjustment and multi-threading—it is now widely supported.
Major hardware such as GPUs~\cite{nguyen2025octopinf}, CPUs~\cite{openvino-batching, demidovskij2020openvino}, TPUs, and NPUs~\cite{dorrich2023tpu}, as well as inference engines like TensorRT~\cite{nguyen2025octopinf, nigade2022jellyfish}, OpenVINO~\cite{openvino-batching, demidovskij2020openvino}, and TensorFlow~\cite{tensorflow-batch}, offer native batching support, making it a scalable optimization across diverse environments.

% A key scalability challenge in FL is the impact of heterogeneous devices, which can hinder performance due to straggler devices~\cite{hu2023gitfl}. 
% To address this issue, \systemname{} adopts two established approaches for managing device performance~\cite{liu2020client,tam2024fedhybrid}. 
% These algorithms are designed to manage device failures and network partitions in FL. 
% When nodes rejoin, they can either continue seamlessly in the next round, provided they are still operational, or they will receive the latest global model to restart their process. 

\subsection{Future Work}
Besides further valuation and fine-tuning to evaluate the performance on other real-time systems, we outline directions for future work aimed at advancing the deployment of \systemname{} in increasingly complex and dynamic real-time edge scenarios.

\textbf{Decentralized and Hybrid Federated Learning (FL).} 
We aim to enable even greater scalability through decentralized or hybrid (FL), where the aggregation of \iagent{}s could occur within networking hardware or on edge devices.
While this concept has been explored in prior work, it has not been considered for real-time inference.
Additionally, further investigation is needed to fully understand the trade-offs and overheads associated with such decentralized aggregation strategies.

\textbf{Global Reward Learning.} 
As application, hardware, and software heterogeneity grows, designing the reward function through manual hyperparameter tuning (see Appendix B) may become increasingly impractical.
To address this, we plan to integrate fast local optimization with global reward learning.
This approach allows the reward function to self-adapt, reducing manual effort while potentially achieving better performance than hand-designed optimization functions.

\subsection{Threats to Validity}

% \textbf{Hardware.} By relying on TensorRT and Nvidia devices, certain performance trade-offs specific to this ecosystem are introduced into the results.
% While alternative hardware or frameworks might yield different outcomes, establishing an alternative evaluation system would be prohibitively expensive.
% The techniques we implement generally exhibit consistent performance trends across various configurations. However, when combined, they may lead to unexpected negative interactions that could affect overall performance.

\par
\textbf{Experiment scenarios.} Despite our best effort, selected videos and scenarios may not fully capture the real-world variability, which could limit the generalization of our findings due to potential data bias.
To mitigate this issue and strengthen the robustness of our conclusions, we collected 23 4-hour videos in different 3 domains--traffic, building surveillance, and audience analysis--with diverse workload patterns (\autoref{fig:batching-dynamics}).
We also incorporated the 2022 AI City Challenge Dataset \cite{Naphade22AIC22}, particularly Track 1: Multi-Camera Vehicle Tracking, to challenge \iagent{}'s learning mechanism on \textit{out-of-distribution} data.
Experimental results from \autoref{subsubsec:eval_warm_start} show that \iagent{} can swiftly adapt to the new dataset.

\par
\textbf{Multiple treatment.} Another challenge we face is the risk of multiple treatment interference, as several changes have been introduced—some of which are fundamental.%, such as shifting the agent from a device-wide approach to a task-specific one.
This makes it difficult to isolate the effects of each individual adjustment.
To address this, we conduct ablation studies to evaluate each proposed technique separately.
The results indicate that each method contributes positively to overall performance, although accurately quantifying the specific impact remains challenging.

% \subsection{Data Availability}
% \label{sub:data}

% The data and code supporting the findings of this study are partially available. 
% The implementation used for the evaluations is provided as an open-source project on GitHub, where you can also find detailed documentation to help reproduce the experimental results. 
% Moreover, this code is archived on Zenodo for long-term preservation. 
% %The implementation used for the evaluations is provided as an open-source project on GitHub\footnote{\href{\sourcecodelink}{\sourcecodelink}}, where you can also find detailed documentation to help reproduce the experimental results. 
% %Moreover, this code is archived on Zenodo\footnote{\href{\archivelink}{\archivelink}} for long-term preservation. 
% However, due to ownership and data privacy concerns, the underlying data cannot be made publicly available. 
% Interested parties can still recreate nearly identical datasets using the provided scripts.

\section{Related Work}
\label{sec:related-work}

\textbf{Edge Workload Distribution.}
For real-time inference, Zhao et al.\cite{zhao2021edgeml} propose a reinforcement learning-based scheme that enhances throughput by selectively offloading parts of a model that cannot fully run on-device.
Works like~\cite{zeng2020distream, Hou2023dystri, gao2024energy, nguyen2025octopinf} focus on balancing VA pipelines between edge devices and servers, by considering network conditions, workload changes, or energy consumption.
CoEdge~\cite{jiang2023coedge}, combines DNN performance estimation with an inference scheduler for containerized co-execution and batching.
Guan et al.\cite{guan2024mixed} propose federated scheduling for DAG Tasks as in \autoref{fig:pipelines} in mixed-critically systems.
As solutions differ in resource centralization, execution flexibility, and throughput gains; self-learning approaches like \systemname{} can adapt accordingly. 

\textbf{Real-Time Reinforcement Learning (RL).} 
Many areas like robotics rely on real-time on-device RL for modern applications under resource constraints~\cite{li2023mathrm}. 
To improve the performance of these applications, Liu et al.\cite{liu2024deadline} present a new formulation for deadline-safe RL execution and Shirvani et al.\cite{shirvani2024duojoule} balance speed, accuracy, and energy dynamically.
However, \systemname{} is too light-weight to benefit from these for significant performance improvements.

\textbf{Edge Federated Learning (FL)} faces unique challenges in heterogeneous edge environments.
Lightweight agents can be achieved using symmetric conversion modules or early exits~\cite{liu2024lightweightFL}.
Zhang et al.\cite{zhang2024ensuring} address bottleneck devices with a 3-stage client selection process, while hierarchical FL clusters agents into groups\cite{zhang2024mobility}.
Asynchronous FL eliminates rounds to handle timing challenges, but suffers from inconsistent and biased updates.
Hu et al.\cite{hu2023gitfl} address this through branch models, which are then aggregated into the global model.
\systemname{} encounters much heterogeneity at runtime, benefiting from advanced FL approaches.

\textbf{Edge Continual Learning (CL).} Deng et al.\cite{deng2023fedinc} propose semi-supervised on-device CL for classification, while Yu et al.\cite{yu2024lifehd} explore lightweight unsupervised clustering using hyperdimensional computing.
Adapting to embedded device constraints, Kwon et al.~\cite{kwon2023lifelearner} present a hardware-aware meta CL system.
These FL and CL advancements highlight their feasibility in Edge environments.
However, applying CL in reinforcement learning, a distinct learning paradigm, remains an open question requiring evaluation through \systemname{}.

\section{Conclusion}
\label{sec:conclusion}

% We present \systemname{}, an optimization system, that combines FRL and CRL to improve the throughput of Edge Visual Analytics inference.
% The dynamics within an environment, posed by unstable network conditions and content variations, offer huge potential for continual configuration adaptation.
% \ifconferenceformat
% Each step within a VA pipeline is optimized locally by an \iagent{} which is dynamically choosing the input resolution, batch size, and multi-thread configuration.

% \systemname{} demonstrates the importance of local optimization in Edge systems.
% We improve the speed and efficiency of machine learning tasks, while also promoting better utilization of resources, leading to more sustainable and cost-effective computing solutions.
% This system signifies a step further towards efficient distributed inference on embedded devices.

In this paper, we present \systemname{}, an optimization system that combines FRL and CRL to enhance the throughput of real-time Edge Visual Analytics (VA) inference.
The dynamics of unstable network conditions and content variations provide significant opportunities for continual configuration adaptation.
Each step in a VA pipeline is locally optimized by an \iagent{}, which dynamically adjusts input resolution, batch size, and multi-thread configuration.
The experimental results show \systemname{} significantly outperforms recent baselines.

\systemname{} highlights the importance of local optimization in real-time edge systems, improving the speed and efficiency of machine learning tasks while ensuring better resource utilization.
This results in more sustainable and cost-effective computing solutions, marking a step forward in efficient real-time distributed inference on embedded devices.

\section*{Reproducibility}
The implementation used for the evaluations is provided as an open-source project on GitHub\footnote{\href{\sourcecodelink}{\sourcecodelink}}, together with detailed documentation to help reproduce the experimental results. 
Moreover, this code is archived on Zenodo\footnote{\href{\archivelink}{\archivelink}} for long-term preservation. 
Though the data is not made publicly available due to ownership and data privacy concerns, interested researchers and developers can still recreate nearly identical datasets for academic purposes with our provided scripts.

\section*{Acknowledgement}
This work was supported by Institute of Information \& communications Technology Planning \& Evaluation (IITP) grant funded by the Korea government (MSIT) (under grant No. RS-2019-II191126, Self-learning based Autonomic IoT Edge Computing), the KAIST Convergence Research Institute Operation Program, and KAIST Deep Mobility Consortium.
The authors acknowledge the use of different AI systems (incl. GPT-4o) limited to editing and grammar enhancement.

\bibliographystyle{IEEEtran}
\bibliography{bibliography}

\begin{thebibliography}{10}
\providecommand{\url}[1]{#1}
\csname url@samestyle\endcsname
\providecommand{\newblock}{\relax}
\providecommand{\bibinfo}[2]{#2}
\providecommand{\BIBentrySTDinterwordspacing}{\spaceskip=0pt\relax}
\providecommand{\BIBentryALTinterwordstretchfactor}{4}
\providecommand{\BIBentryALTinterwordspacing}{\spaceskip=\fontdimen2\font plus
\BIBentryALTinterwordstretchfactor\fontdimen3\font minus
  \fontdimen4\font\relax}
\providecommand{\BIBforeignlanguage}[2]{{%
\expandafter\ifx\csname l@#1\endcsname\relax
\typeout{** WARNING: IEEEtran.bst: No hyphenation pattern has been}%
\typeout{** loaded for the language `#1'. Using the pattern for}%
\typeout{** the default language instead.}%
\else
\language=\csname l@#1\endcsname
\fi
#2}}
\providecommand{\BIBdecl}{\relax}
\BIBdecl

\bibitem{Ananthanarayanan2017vakiller}
G.~Ananthanarayanan, P.~Bahl, P.~Bodik, K.~Chintalapudi, M.~Philipose,
  L.~Ravindranath, and S.~Sinha, ``{Real-Time Video Analytics: The Killer App
  for Edge Computing},'' \emph{Computer}, vol.~50, no.~10, pp. 58--67, 2017.

\bibitem{Nguyen2023preacto}
T.-T. Nguyen, S.~Y. Jang, B.~Kostadinov, and D.~Lee, ``{PreActo: Efficient
  Cross-Camera Object Tracking System in Video Analytics Edge Computing},'' in
  \emph{2023 IEEE Int. Conf. on Pervasive Computing and Communications
  (PerCom)}.\hskip 1em plus 0.5em minus 0.4em\relax IEEE, 2023, pp. 101--110.

\bibitem{Hung2018VideoEdge}
C.-C. Hung, G.~Ananthanarayanan, P.~Bodik, L.~Golubchik, M.~Yu, P.~Bahl, and
  M.~Philipose, ``Videoedge: Processing camera streams using hierarchical
  clusters,'' in \emph{2018 IEEE/ACM Symposium on Edge Computing (SEC)}, 2018,
  pp. 115--131.

\bibitem{nguyen2025octopinf}
T.-T. Nguyen, L.~Liebe, T.-N. Quang, Y.~Wu, J.~Cheng, and D.~Lee, ``{OCTOPINF:
  Workload-Aware Real-Time Inference Serving for Edge Video Analytics},'' in
  \emph{The 23rd International Conference on Pervasive Computing and
  Communications (PerCom 2025)}.\hskip 1em plus 0.5em minus 0.4em\relax IEEE,
  2025.

\bibitem{zeng2020distream}
X.~Zeng, B.~Fang, H.~Shen, and M.~Zhang, ``Distream: scaling live video
  analytics with workload-adaptive distributed edge intelligence,'' in
  \emph{Procs. of the 18th SenSys}, 2020, pp. 409--421.

\bibitem{Hou2023dystri}
X.~Hou, Y.~Guan, and T.~Han, ``{Dystri: A Dynamic Inference based Distributed
  DNN Service Framework on Edge},'' in \emph{ACM Int. Conf. Proceeding
  Series}.\hskip 1em plus 0.5em minus 0.4em\relax ACM, aug 2023, pp. 625--634.

\bibitem{ma2024performance}
Y.~Ma, R.~Fu, A.~Zou, J.~Li, C.~Chen, C.~Lu, and X.~Guan, ``Performance
  optimization and stability guarantees for multi-tier real-time control
  systems,'' in \emph{2024 IEEE Real-Time Systems Symposium (RTSS)}.\hskip 1em
  plus 0.5em minus 0.4em\relax IEEE, 2024, pp. 187--200.

\bibitem{gao2024energy}
C.~Gao, N.~Kumar, and A.~Easwaran, ``Energy-efficient real-time job mapping and
  resource management in mobile-edge computing,'' in \emph{2024 IEEE Real-Time
  Systems Symposium (RTSS)}.\hskip 1em plus 0.5em minus 0.4em\relax IEEE, 2024,
  pp. 15--28.

\bibitem{chen2024scenic}
J.~Chen, A.~Zou, Y.~Xu, and Y.~Ma, ``Scenic: Capability and scheduling
  co-design for intelligent controller on heterogeneous platforms,'' in
  \emph{2024 IEEE Real-Time Systems Symposium (RTSS)}.\hskip 1em plus 0.5em
  minus 0.4em\relax IEEE, 2024, pp. 1--14.

\bibitem{zhang2024bcedge}
Z.~Zhang, Y.~Zhao, H.~Li, and J.~Liu, ``Bcedge: Slo-aware dnn inference
  services with adaptive batch-concurrent scheduling on edge devices,''
  \emph{IEEE Transactions on Network and Service Management}, 2024.

\bibitem{fang2017qos}
Z.~Fang, T.~Yu, O.~J. Mengshoel, and R.~K. Gupta, ``Qos-aware scheduling of
  heterogeneous servers for inference in deep neural networks,'' in
  \emph{Procs. of the 2017 ACM on Conf. on Information and Knowledge
  Management}, 2017, pp. 2067--2070.

\bibitem{she2024earlyexit}
Y.~She, T.~Shi, J.~Wang, and B.~Liu, ``Dynamic batching and early-exiting for
  accurate and timely edge inference,'' in \emph{2024 IEEE 99th Vehicular
  Technology Conf. (VTC2024-Spring)}, 2024.

\bibitem{coviello2021magicpipe}
G.~Coviello, Y.~Yang, K.~Rao, and S.~Chakradhar, ``Magic-pipe: Self-optimizing
  video analytics pipelines,'' in \emph{Procs. of the 22nd Int. Middleware
  Conference}, 2021, pp. 79--90.

\bibitem{osinenko2023actor}
P.~Osinenko, G.~Yaremenko, G.~Malaniya, and A.~Bolychev, ``An actor-critic
  framework for online control with environment stability guarantee,''
  \emph{IEEE Access}, vol.~11, pp. 89\,188--89\,204, 2023.

\bibitem{abel2024continual}
D.~Abel, A.~Barreto, B.~Van~Roy, D.~Precup, H.~P. van Hasselt, and S.~Singh,
  ``A definition of continual reinforcement learning,'' \emph{Advances in
  Neural Information Processing Systems}, vol.~36, 2024.

\bibitem{Jin2022federatedreinforcementlearning}
H.~Jin, Y.~Peng, W.~Yang, S.~Wang, and Z.~Zhang, ``{Federated Reinforcement
  Learning with Environment Heterogeneity},'' \emph{Procs. of Machine Learning
  Research}, vol. 151, pp. 18--37, 2022.

\bibitem{ali2020batch}
A.~Ali, R.~Pinciroli, F.~Yan, and E.~Smirni, ``Batch: Machine learning
  inference serving on serverless platforms with adaptive batching,'' in
  \emph{SC20: Int. Conf. for High Performance Computing, Networking, Storage
  and Analysis}.\hskip 1em plus 0.5em minus 0.4em\relax IEEE, 2020, pp. 1--15.

\bibitem{nigade2022jellyfish}
V.~Nigade, P.~Bauszat, H.~Bal, and L.~Wang, ``Jellyfish: Timely inference
  serving for dynamic edge networks,'' in \emph{2022 IEEE Real-Time Systems
  Symposium (RTSS)}.\hskip 1em plus 0.5em minus 0.4em\relax IEEE, 2022, pp.
  277--290.

\bibitem{gokarn2023mosaic}
I.~Gokarn, H.~Sabbella, Y.~Hu, T.~Abdelzaher, and A.~Misra, ``Mosaic:
  Spatially-multiplexed edge ai optimization over multiple concurrent video
  sensing streams,'' in \emph{Procs. of the 14th Conf. on ACM Multimedia
  Systems}, 2023, pp. 278--288.

\bibitem{peng2024tangram}
H.~Peng, Y.~Zhan, P.~Li, and Y.~Xia, ``Tangram: High-resolution video analytics
  on serverless platform with slo-aware batching,'' in \emph{2024 IEEE 44th
  Int. Conf. on Distributed Computing Systems (ICDCS)}.\hskip 1em plus 0.5em
  minus 0.4em\relax IEEE, 2024, pp. 645--655.

\bibitem{mansour2021rcnn}
R.~F. Mansour, J.~Escorcia-Gutierrez, M.~Gamarra, J.~A. Villanueva, and
  N.~Leal, ``Intelligent video anomaly detection and classification using
  faster rcnn with deep reinforcement learning model,'' \emph{Image and Vision
  Computing}, vol. 112, p. 104229, 2021.

\bibitem{Lubben2011networkmdp}
R.~Lubben, M.~Fidler, and J.~Liebeherr, ``{A foundation for stochastic
  bandwidth estimation of networks with random service},'' in \emph{2011
  Proceedings IEEE INFOCOM}.\hskip 1em plus 0.5em minus 0.4em\relax IEEE, apr
  2011, pp. 1817--1825.

\bibitem{schulman2017proximal}
J.~Schulman, F.~Wolski, P.~Dhariwal, A.~Radford, and O.~Klimov, ``Proximal
  policy optimization algorithms,'' \emph{arXiv:1707.06347}, 2017.

\bibitem{tam2024fedhybrid}
K.~Tam, C.~Tian, L.~Li, H.~Zhao, and C.~Xu, ``Fedhybrid: Breaking the memory
  wall of federated learning via hybrid tensor management,'' in \emph{Procs. of
  the 22nd SenSys}, 2024, pp. 394--408.

\bibitem{liu2020client}
L.~Liu, J.~Zhang, S.~Song, and K.~B. Letaief, ``Client-edge-cloud hierarchical
  federated learning,'' in \emph{ICC 2020-2020 IEEE international conference on
  communications (ICC)}.\hskip 1em plus 0.5em minus 0.4em\relax IEEE, 2020, pp.
  1--6.

\bibitem{raca2020beyond}
D.~Raca, D.~Leahy, C.~J. Sreenan, and J.~J. Quinlan, ``Beyond throughput, the
  next generation: A 5g dataset with channel and context metrics,'' in
  \emph{Procs. of the 11th ACM multimedia systems Conf.}, 2020, pp. 303--308.

\bibitem{pipelinescheduler}
\BIBentryALTinterwordspacing
T.-T. Nguyen, L.~Liebe, T.~N. Quang, Y.~Wu, chengjinghan, and cdsnlab,
  ``tungngreen/pipelinescheduler: v0.1.0 - percom 2025 implementation,'' Feb.
  2025. [Online]. Available: \url{https://doi.org/10.5281/zenodo.14789255}
\BIBentrySTDinterwordspacing

\bibitem{Naphade22AIC22}
M.~Naphade, S.~Wang, D.~C. Anastasiu, Z.~Tang, M.~Chang, Y.~Yao, L.~Zheng,
  M.~S. Rahman, A.~Venkatachalapathy, A.~Sharma, Q.~Feng, V.~Ablavsky,
  S.~Sclaroff, P.~Chakraborty, A.~Li, S.~Li, and R.~Chellappa, ``The 6th ai
  city challenge,'' in \emph{2022 IEEE/CVF Conference on Computer Vision and
  Pattern Recognition Workshops (CVPRW)}.\hskip 1em plus 0.5em minus
  0.4em\relax IEEE Computer Society, June 2022, pp. 3346--3355.

\bibitem{hayek2025federated}
R.~J. Hayek, J.~Chung, K.~Comer, C.~R. Murthy, R.~Kettimuthu, and I.~Kadota,
  ``Federated learning over 5g, wifi, and ethernet: Measurements and
  evaluation,'' \emph{arXiv preprint arXiv:2504.04678}, 2025.

\bibitem{mcmahan2017communication}
B.~McMahan, E.~Moore, D.~Ramage, S.~Hampson, and B.~A. y~Arcas,
  ``Communication-efficient learning of deep networks from decentralized
  data,'' in \emph{Artificial intelligence and statistics}.\hskip 1em plus
  0.5em minus 0.4em\relax PMLR, 2017, pp. 1273--1282.

\bibitem{gao2020batchrnn}
\BIBentryALTinterwordspacing
C.~Gao, A.~Rios-Navarro, X.~Chen, S.-C. Liu, and T.~Delbruck, ``{EdgeDRNN}:
  {Recurrent} {Neural} {Network} {Accelerator} for {Edge} {Inference},''
  \emph{IEEE Journal on Emerging and Selected Topics in Circuits and Systems},
  vol.~10, no.~4, pp. 419--432, Feb. 2020. [Online]. Available:
  \url{https://ieeexplore.ieee.org/abstract/document/9268992}
\BIBentrySTDinterwordspacing

\bibitem{holmes2019batchrnn}
\BIBentryALTinterwordspacing
C.~Holmes, D.~Mawhirter, Y.~He, F.~Yan, and B.~Wu,
  ``\BIBforeignlanguage{en}{{GRNN}: {Low}-{Latency} and {Scalable} {RNN}
  {Inference} on {GPUs}},'' in \emph{\BIBforeignlanguage{en}{Proceedings of the
  {Fourteenth} {EuroSys} {Conference} 2019}}.\hskip 1em plus 0.5em minus
  0.4em\relax Dresden Germany: ACM, Mar. 2019, pp. 1--16. [Online]. Available:
  \url{https://dl.acm.org/doi/10.1145/3302424.3303949}
\BIBentrySTDinterwordspacing

\bibitem{noauthor2025batchtransformer}
\BIBentryALTinterwordspacing
``{NVIDIA}/{FasterTransformer},'' May 2025, original-date:
  2021-04-02T21:36:33Z. [Online]. Available:
  \url{https://github.com/NVIDIA/FasterTransformer}
\BIBentrySTDinterwordspacing

\bibitem{noauthor2023batchllm}
\BIBentryALTinterwordspacing
``\BIBforeignlanguage{en-US}{Mastering {LLM} {Techniques}: {Inference}
  {Optimization}},'' Nov. 2023. [Online]. Available:
  \url{https://developer.nvidia.com/blog/mastering-llm-techniques-inference-optimization/}
\BIBentrySTDinterwordspacing

\bibitem{openvino-batching}
\BIBentryALTinterwordspacing
``Automatic {Batching} — {OpenVINO}™ {documentationCopy} to {clipboardCopy}
  to {clipboardCopy} to {clipboardCopy} to {clipboardCopy} to {clipboardCopy}
  to {clipboardCopy} to {clipboardCopy} to {clipboardCopy} to {clipboardCopy}
  to {clipboardCopy} to {clipboardCopy} to clipboard — {Version}(2024).''
  [Online]. Available:
  \url{https://docs.openvino.ai/2024/openvino-workflow/running-inference/inference-devices-and-modes/automatic-batching.html}
\BIBentrySTDinterwordspacing

\bibitem{demidovskij2020openvino}
\BIBentryALTinterwordspacing
A.~Demidovskij, A.~Tugaryov, A.~Suvorov, Y.~Tarkan, M.~Fatekhov, I.~Salnikov,
  A.~Kashchikhin, V.~Golubenko, G.~Dedyukhina, A.~Alborova, R.~Palmer,
  M.~Fedorov, and Y.~Gorbachev, ``{OpenVINO} {Deep} {Learning} {Workbench}: {A}
  {Platform} for {Model} {Optimization}, {Analysis} and {Deployment},'' in
  \emph{2020 {IEEE} 32nd {International} {Conference} on {Tools} with
  {Artificial} {Intelligence} ({ICTAI})}, Jan. 2020, pp. 661--668, iSSN:
  2375-0197. [Online]. Available:
  \url{https://ieeexplore.ieee.org/document/9288163}
\BIBentrySTDinterwordspacing

\bibitem{dorrich2023tpu}
\BIBentryALTinterwordspacing
M.~Dörrich, M.~Fan, and A.~M. Kist, ``Impact of {Mixed} {Precision}
  {Techniques} on {Training} and {Inference} {Efficiency} of {Deep} {Neural}
  {Networks},'' \emph{IEEE Access}, vol.~11, pp. 57\,627--57\,634, 2023.
  [Online]. Available: \url{https://ieeexplore.ieee.org/document/10146255}
\BIBentrySTDinterwordspacing

\bibitem{tensorflow-batch}
\BIBentryALTinterwordspacing
``\BIBforeignlanguage{en}{Tensorflow {Serving} {Configuration} {\textbar}
  {TFX}}.'' [Online]. Available:
  \url{https://www.tensorflow.org/tfx/serving/serving_config}
\BIBentrySTDinterwordspacing

\bibitem{zhao2021edgeml}
Z.~Zhao, K.~Wang, N.~Ling, and G.~Xing, ``Edgeml: An automl framework for
  real-time deep learning on the edge,'' in \emph{Procs. of IoTDI}, 2021, pp.
  133--144.

\bibitem{jiang2023coedge}
Z.~Jiang, N.~Ling, X.~Huang, S.~Shi, C.~Wu, X.~Zhao, Z.~Yan, and G.~Xing,
  ``Coedge: A cooperative edge system for distributed real-time deep learning
  tasks,'' in \emph{Procs. of the 22nd Int. Conf. on Information Processing in
  Sensor Networks}, 2023, pp. 53--66.

\bibitem{guan2024mixed}
F.~Guan, J.~Lee, C.~J. Xue, J.-M. Wu, and N.~Guan, ``Mixed-criticality
  federated scheduling for relaxed-deadline dag tasks,'' in \emph{2024 IEEE
  Real-Time Systems Symposium (RTSS)}.\hskip 1em plus 0.5em minus 0.4em\relax
  IEEE, 2024, pp. 362--374.

\bibitem{li2023mathrm}
Z.~Li, A.~Samanta, Y.~Li, A.~Soltoggio, H.~Kim, and C.~Liu,
  ``$\backslash$mathrm $\{$R$\}$\^{}$\{$3$\}$: On-device real-time deep
  reinforcement learning for autonomous robotics,'' in \emph{2023 IEEE
  Real-Time Systems Symposium (RTSS)}.\hskip 1em plus 0.5em minus 0.4em\relax
  IEEE, 2023, pp. 131--144.

\bibitem{liu2024deadline}
M.~Liu, P.~Lu, X.~Chen, O.~Sokolsky, I.~Lee, and F.~Kong, ``Deadline-safe
  reach-avoid control synthesis for cyber-physical systems with reinforcement
  learning,'' in \emph{2024 IEEE Real-Time Systems Symposium (RTSS)}.\hskip 1em
  plus 0.5em minus 0.4em\relax IEEE, 2024, pp. 96--108.

\bibitem{shirvani2024duojoule}
S.~Shirvani, A.~Samanta, Z.~Li, and C.~Liu, ``Duojoule: Accurate on-device deep
  reinforcement learning for energy and timeliness,'' in \emph{2024 IEEE
  Real-Time Systems Symposium (RTSS)}.\hskip 1em plus 0.5em minus 0.4em\relax
  IEEE, 2024, pp. 109--122.

\bibitem{liu2024lightweightFL}
J.~Liu, H.~Huang, C.~Wang, R.~Li, T.~Car, Q.~Yang, and Z.~Zheng, ``Can
  federated learning clients be lightweight? a plug-and-play symmetric
  conversion module,'' in \emph{2024 IEEE 44th Int. Conf. on Distributed
  Computing Systems (ICDCS)}.\hskip 1em plus 0.5em minus 0.4em\relax IEEE,
  2024, pp. 809--820.

\bibitem{zhang2024ensuring}
M.~Zhang, H.~Zhao, S.~Ebron, R.~Xie, and K.~Yang, ``Ensuring fairness in
  federated learning services: Innovative approaches to client selection,
  scheduling, and rewards,'' in \emph{2024 IEEE 44th Int. Conf. on Distributed
  Computing Systems (ICDCS)}.\hskip 1em plus 0.5em minus 0.4em\relax IEEE,
  2024, pp. 762--773.

\bibitem{zhang2024mobility}
S.~Zhang, Z.~Zheng, Q.~Li, F.~Wu, and G.~Chen, ``Mobility-aware device sampling
  for statistical heterogeneity in hierarchical federated learning,'' in
  \emph{2024 IEEE 44th Int. Conf. on Distributed Computing Systems
  (ICDCS)}.\hskip 1em plus 0.5em minus 0.4em\relax IEEE, 2024, pp. 656--667.

\bibitem{hu2023gitfl}
M.~Hu, Z.~Xia, D.~Yan, Z.~Yue, J.~Xia, Y.~Huang, Y.~Liu, and M.~Chen, ``Gitfl:
  Uncertainty-aware real-time asynchronous federated learning using version
  control,'' in \emph{2023 IEEE Real-Time Systems Symposium (RTSS)}.\hskip 1em
  plus 0.5em minus 0.4em\relax IEEE, 2023, pp. 145--157.

\bibitem{deng2023fedinc}
Y.~Deng, S.~Yue, T.~Wang, G.~Wang, J.~Ren, and Y.~Zhang, ``Fedinc: An
  exemplar-free continual federated learning framework with small labeled
  data,'' in \emph{Procs. of the 21st ACM SenSys}, 2023, pp. 56--69.

\bibitem{yu2024lifehd}
X.~Yu, A.~Thomas, I.~G. Moreno, L.~Gutierrez, and T.~S. Rosing, ``Intelligence
  beyond the edge using hyperdimensional computing,'' in \emph{2024 23rd
  ACM/IEEE Int. Conf. on Information Processing in Sensor Networks (IPSN)},
  2024, pp. 1--13.

\bibitem{kwon2023lifelearner}
Y.~D. Kwon, J.~Chauhan, H.~Jia, S.~I. Venieris, and C.~Mascolo, ``Lifelearner:
  Hardware-aware meta continual learning system for embedded computing
  platforms,'' in \emph{Procs. of the 21st ACM Conf. on Embedded Networked
  Sensor Systems}, 2023, pp. 138--151.

\end{thebibliography}

\end{document}